\documentstyle{article}
\input boxedeps.tex
\SetRokickiEPSFSpecial  
\HideDisplacementBoxes
\newcommand{\da}{\dagger}  
\newcommand{\be}{\begin{equation}}
\newcommand{\eq}{\end{equation}}
\newcommand{\Tr}{{\rm \, Tr}}
\newcommand{\LCTL}{LCTL}

\begin{document}
\mbox{}\hfill DAMTP-97-16\\
\mbox{}\hfill hep-ph/9704408\\
\vspace{10mm}
\begin{center}
{\LARGE Colour-Dielectric Gauge Theory\\
on a Transverse Lattice\\}
\vspace{30mm}

{\bf S. Dalley}${}^{*}$\footnote{From 1st April 1997 on leave at: 
Theory Division, CERN, CH-1211
Geneva 23, Switzerland.} and  {\bf B. van de Sande}${}^{**}$\\
\vspace{10mm}

{\em
${}^*$Department of Applied Mathematics and Theoretical Physics\\
Silver Street, Cambridge CB3 9EW, England\\
\vspace{5mm}

${}^{**}$Institut F\"ur Theoretische Physik III,\\
Staudstra{\ss}e 7, D-91058 Erlangen, Germany }
\vspace{30mm}

\end{center} 

\begin{abstract}
We investigate in some detail consequences of the
effective colour-dielectric formulation of lattice gauge theory
using the light-cone Hamiltonian formalism with a transverse
lattice. As a quantitative test of this approach, 
we have performed extensive analytic
and numerical calculations for $2+1$-dimensional pure gauge theory
in the large $N$ limit. Because of  Eguchi-Kawai reduction, one
effectively studies a $1+1$-dimensional gauge theory coupled to
matter in the adjoint representation. 
We study the structure of coupling constant space for our
effective potential by comparing with 
results available from  
conventional Euclidean lattice Monte Carlo simulations of this system. 
In particular, we calculate and measure the
scaling behaviour of the entire low-lying glueball spectrum, glueball
wavefunctions, string tension, asymptotic density of states, and deconfining
temperature. 
We employ a new hybrid DLCQ/wavefunction basis 
in our calculations of the light-cone Hamiltonian matrix elements,
along with extrapolation in Tamm-Dancoff truncation, 
significantly reducing numerical errors. Finally we discuss, in
light of our results, what further measurements and calculations
could be made in order to systematically remove lattice spacing
dependence from our effective potential {\em a priori}.

\end{abstract}

\newpage
\baselineskip .25in
\section{Introduction}
The least understood scale of the strong interactions is the
intermediate one, between short distances described by asymptotically
free QCD and the relatively long range phenomena of
nuclear forces. A successful description
of this intermediate region should not only include `global' properties
of hadrons, such as masses and decay constants,
but also  the wealth of accumulated
experimental data on hadronic sub-structure.
In addition one might hope to
shed some light on the interface between gauge theory and nuclear physics.
There are also many reasons why knowledge of hadron wavefunctions in
a general Lorentz frame, particularly at the
amplitude level, will be necessary for future progress in all
aspects of particle physics. 
At present, no single approach to the strong interactions seems 
capable of this. In this paper we pursue an idea which, in principle,
can address the above problems. As we will demonstrate, 
the colour-dielectric concept is both a natural description and, when 
combined with the light-cone Hamiltonian framework,  
a practical description of the strong interactions at intermediate scales.  
A number of authors have
proposed the colour dielectric idea in various forms 
\cite{bard}--\cite{adler}\footnote{A review of subsequent
investigations and more extensive references can be found in 
Ref.~\cite{pirner}.} but, while the motivation was often
elegant, results were usually of a qualitative nature.
Our work is strongly foreshadowed by the
work of Bardeen, Pearson, and Rabinovici \cite{bard}
in which the `dielectric' formulation of a light-cone transverse lattice
gauge theory was first proposed (although the authors did not discuss
it in quite those terms).  

We characterise a dielectric formulation as one in which gluon fields,
or rather the $SU(N)$ group elements they generate, are replaced by 
collective variables which represent an average over the 
fluctuations on short distance scales.
These dielectric variables carry colour and form an effective gauge
field theory
with classical action minimised at zero field, 
meaning that colour flux  is expelled from the vacuum at the classical level. 
The price one pays for starting with a simple vacuum structure,
which may arise only for a rather low momentum cut-off on the effective theory,
is that the effective action will be largely unknown and must be 
investigated {\em per se}. `Conservation of
difficulty' appears to have dashed one's hopes. However, there are at
least two reasons why shifting the complication from the vacuum to the
action is desirable. Firstly, although most approaches to field theory
place great emphasis on the calculation of vacuum energy and field
condensates, neither of these are  measured in experiment. Secondly,
if one hopes to employ light-cone quantisation, which seems the only
practical way to calculate the hadron's wavefunction in a general
Lorentz frame, then dealing with explicit vacuum structure can be rather
awkward.

We choose to employ a lattice dielectric gauge theory in which the
fields are $N$ by $N$ complex matrices $M$, in the adjoint
representation of the gauge group, attached to links of the 
lattice \cite{mack}. 
Because the lattice spacing in this
theory should be rather large if the colour dielectric idea is to be valid,
it is essentially impossible to obtain the effective action by
explicitly
integrating out gluons (blocking from the continuum). More realistic,
but still essentially impossible, would be to introduce the dielectric
fields by blocking the unitary link fields $U \in SU(N)$ of Wilson's lattice
gauge theory \cite{wilson}. 
That leaves a `trial-and-error' approach where one tunes the
coefficients of the effective action, or some truncation of it, until
solutions exhibit the correct symmetries, such as Lorentz invariance.
This more-or-less rules out a Euclidean path integral formulation, since one
would have to do the integral (numerically) for each set of
coefficients. 
A Hamiltonian approach is a viable alternative since most
of the work goes into computing matrix elements of the Hamiltonian. 
This only needs to be done once for each operator appearing in the
effective action. It is only 
the diagonalisation of the Hamiltonian, which can be
performed much more rapidly, that needs to be repeated for different
coefficients of the operators.
In the light-cone Hamiltonian formulation \cite{bard} 
the continuum limit is taken in the temporal and in one
spatial direction, leaving a transverse lattice.
By using  a light-cone Hamiltonian formulation 
the quantum vacuum remains as simple (modulo zero modes)
as the classical vacuum. 

Next we must choose a specific problem in order to test the
practicality of this dielectric formulation. In this paper we 
postpone the introduction fermions 
(and hence quark structure functions) and study
pure gauge theory. We look at glueballs in three spacetime dimensions at
large $N$. The glueball problem is one of the most difficult
one could imagine since it is ultra-relativistic and strongly
interacting. However in {\em four} dimensions there is a dearth of accurate,
detailed experimental or theoretical results which could be used as 
a benchmark. While there is an undoubted urge to contribute to the presently
controversial debate on glueballs in nature, as a first calculation
a  more circumspect choice is appropriate. Fortunately there is a 
wealth of accurate information now available on glueballs 
for pure gauge theory in {\em three} dimensions from the recent
Euclidean Lattice Monte Carlo (ELMC) simulations of  Teper
\cite{teper}. 
That work has shown that pure non-Abelian
gauge theory behaves much the same way in three as in four dimensions:
a discrete set of massive boundstates are  generated by
a linearly confining string-like force.
Moreover, Teper has performed calculations for $2$, $3$, and $4$
colours, allowing an extrapolation to the large $N$ limit. The relevant 
expansion parameter is $1/N^2$ and observables measured at $1/N^2 = 0$
are extremely close to those measured at $1/N^2 = 1/9$. The large-$N$
limit  is convenient, though not essential,
for our Light-Cone Transverse Lattice (\LCTL{}) formulation, 
which becomes simpler
in the limit of many colours. It enables us to implement Eguchi-Kawai
reduction \cite{ek} to a $1+1$-dimensional theory 
which does not need to be quenched or twisted
so long as the  vacuum is the simple dielectric one.
It also allows us to make use of the extensive intuition developed
for large-$N$ theories in relation to string models. Indeed, our
formulation offers the rare possibility of describing the parton,
constituent, and string behaviour of hadrons in one framework. The 
relationship between these pictures, each very different but equally
successful, remains one of the outstanding enigmas of QCD.

We should emphasize that this is the first time that the \LCTL{}
formulation has been used in a serious quantitative study, 
and the expectations of
the reader should be tempered accordingly. It was necessary to
overcome a number of technical problems whose solutions
probably have wider applicability.
Our initial goal has been to explore the coupling constant space of
the dielectric effective potential, which was largely ignored by Bardeen 
{\em et alii} \cite{bard} in their pioneering study.
We will fix our effective potential by comparison with the the ELMC spectrum. 
This gives rise to predictions for other observables
and it is non-trivial that a consistent picture emerges.
In the light of our results, we will also discuss ways of measuring
further observables to test Lorentz covariance and so fix the
effective potential independently, as well as the {\em sine qua non} 
of a systematically improvable analytic renormalisation group analysis.

The outline of the paper is as follows. In the following section
we describe in general terms the dielectric formulation 
on a Euclidean lattice, with particular reference to the large $N$ limit. 
The associated light-cone Hamiltonian limit of the transverse lattice 
is constructed in Section~\ref{sec3}. 
In Section~\ref{sec4} we obtain analytic and numerical solutions for glueballs,
compare with Teper's data \cite{teper}, and
identify a scaling trajectory in coupling constant space. Section~\ref{sec5}
addresses the string tension via winding modes.
We discuss the asymptotic density of glueball states and associated finite 
temperature properties in Section~\ref{sec6} and compare with 
results from string theory. 
The conclusions in Section~\ref{conclude} 
contain discussion of  some general points of principle, as well
as the shortcomings of \LCTL{} as used
in this paper and their possible resolution.
Appendix~\ref{appendixa}
 contains some details of the proof of Eguchi-Kawai reduction
in light-cone Hamiltonian formalism. In order to obtain the
required numerical accuracy, we had to devise new methods of computation
and extrapolation for light-cone Hamiltonians. These important
techniques are collected in  Appendices B and C since they are relevant
to many parts of the paper.

Some of our preliminary results have been presented in Refs.~\cite{us}.

\section{Dielectric gauge theory on a Euclidean lattice}
\label{motivate}

In this section we motivate the colour dielectric formulation of
gauge theory 
in the context of a Euclidean lattice \cite{mack}. 
For intermediate scales, we  stress that we have not
derived it in detail from  QCD at short distances, but 
view it as a starting point, similar to a  sigma model for mesons.  
However, unlike  a  sigma model, or even a quark effective theory,
there will be explicit colour-field degrees of
freedom. The best we can do at this stage is to
discuss the general form expected of the effective potential. The 
details for the transverse lattice will be determined later by 
actually solving the theory as a
function of the associated coupling constants.

\subsection{Effective action}

Let us consider 
Wilson's formulation \cite{wilson} of lattice gauge theory.
The principle observation is that,
for large lattice spacing, the unitary link variables $U$
have large fluctuations away from the identity 
and it may be just as well to replace them by some other set of
variables which takes advantage of this randomness. Nielsen and 
Patkos \cite{niel} chose a `block' transformation from the product
of parallel transporters $U$ along some 
set of lattice paths ${\cal C}_{ab}$ weighted by $\rho ({\cal
C}_{ab})$
between two points $a$ and $b$ of a sub-lattice
\be
M (a,b) = \sum_{\{{\cal C}_{ab}\}} \rho ({\cal C}_{ab})  \prod_{{\cal
C}_{ab}} U \; . \label{block}
\eq
The blocked matrix link variable $M \in GL(N,C)$,  since it is
a linear combination of fairly random unitary variables. This blocking
scheme has been investigated  by operator matching using Monte Carlo and
Schwinger-Dyson techniques \cite{pirner} for $SU(2)$, with conclusions
for the general form of the effective potential similar to 
those described below for large $N$.
Although we will not use the reasoning (\ref{block}) 
directly,
it motivates us to search for
an effective lattice gauge theory in terms of such non-compact link
variables. 

One knows that at large lattice spacing $a$  Wilson's action 
gives a fair description of non-Abelian physics.
One assumes that as $a \to 0$ the lattice errors gradually reduce
to nil.
In order to reduce the lattice errors at a finite $a$
one might imagine `smearing' the
link variables $U$. Therefore consider introducing both a unitary
matrix $U(l)$ and a complex matrix $M(l)$ on each link $l$ of the 
spacetime lattice, with partition function
\be
Z = \int \prod_{l} {\cal D}U(l)\, {\cal D}M(l)\, \exp \left(
{1 \over g^2_{L}} 
\sum_{P} \Tr_P  [M] - {N \over \lambda} \sum_{l} \Tr\left\{(M(l) -
U(l))(M^{\dagger}(l) - U^{\dagger}(l))\right\} \right) \label{smear}
\eq
where $\Tr_{P}[M]$ denotes the trace of the product of link 
variables $M$ around an 
elementary plaquette $P$ and $g_L$ is the dimensionless lattice gauge 
coupling constant. ${\cal D}$ denotes the usual Haar measure,
which is simply
\be
{\cal D}M = \prod_{i,j =1}^{N} dM_{ij} \, dM^*_{ij}
\eq
for the complex matrices.
In this case we have introduced a simple Gaussian smearing function
which peaks the measure of the complex matrices on unitary
matrices. In the limit
$\lambda \to 0$, one recovers Wilson's theory. It is natural therefore
to associate small $\lambda$ with small $a$.
For small $\lambda$
one can derive a series of corrections to the Wilson action 
$\Tr_{P}[U]$
as an expansion in $\lambda$. Define the complex matrix
\be
C(l) = \sqrt{N \over \lambda} \, \left[M(l) - U(l) \right] \; .
\eq
If $1_P$, $2_P$, $3_P$, and $4_P$ are the four links
that form plaquette $P$,
we have
\be
Z  =  \int \prod_{l} {\cal D}U(l) \,
\exp \left( {1 \over g^2_{L} }  \left(\sum_P 
\Tr_{P} [U] \right) + {1 \over g^2_{L} }\Delta V[U]\right)
\eq
where
\begin{eqnarray}
\exp \left( {1 \over g_{L}^{2}} \Delta V[U] \right)
 & = & \int \prod_l {\cal D}C(l) \,
\exp \left(  - \sum_{l} \Tr \left\{C^{\dagger}(l)C(l)\right\} 
\right. \nonumber \\
& &\;\;\;\;+ {\sqrt{\lambda} \over
g^2_{L} \sqrt{N}} \sum_P  \Tr \left\{
  C(1_P)U(2_P)U(3_P)U(4_P) + U(1_P)C(2_P)U(3_P)U(4_P)  
 \right.  \\
&&\;\;\;\;\left. \left. + U(1_P)U(2_P)C(3_P)U(4_P)+ 
U(1_P)U(2_P)U(3_P)C(4_P) \right\} + O(\lambda)  \right) 
\nonumber
\end{eqnarray}
and we have dropped an irrelevant normalisation constant.
Performing the Gaussian integrals over $C(l)$ one generates an effective action for $U$-variables
consisting of arbitrary Wilson loops: 
%
%
\begin{equation}
\setlength{\arraycolsep}{1mm}
\Tr_{P}[U]+\Delta V[U] =\sum_P \BoxedEPSF{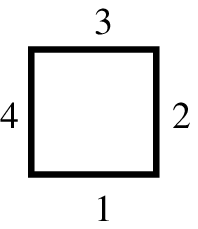 scaled 600}+ 
        \frac{\lambda}{N g_L^2}\left(
           \BoxedEPSF{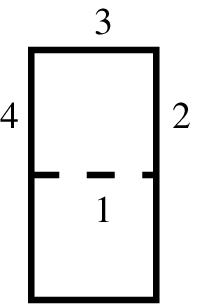 scaled 600} +
           \BoxedEPSF{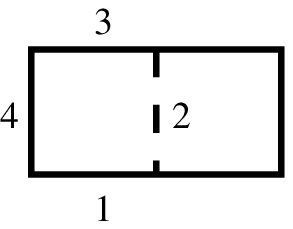 scaled 600} +
           \BoxedEPSF{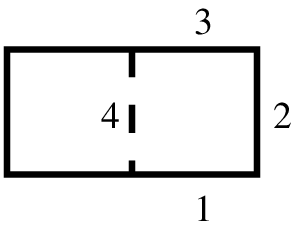 scaled 600} +
           \BoxedEPSF{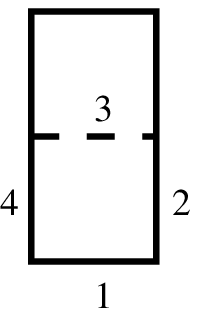 scaled 600}
 \right) + O(\lambda^2) \; . \label{boxes}
\end{equation}
(Only the terms in one plane are shown.)
The leading correction is
the well known rectangular $U$-loop which, with the correct coefficient,
will cancel $O(a^2)$ errors of the Wilson action. One might imagine
that by introducing a sufficiently complicated smearing function one
could produce systematically the terms needed to correct for higher
lattice spacing errors, {\em id est} a perfect action which tracks an 
infrared stable RG trajectory \cite{wilson2}. 
With the simple Gaussian smearing, the
first correction to Wilson's action 
(\ref{boxes}) in fact has the wrong sign for improvement. However we 
are free to make the effective potential more sophisticated beyond the
Gaussian approximation by explicitly adding 
zero-area Wilson $M$-loops (which become trivial in the
limit $\lambda \to 0$) to the action (\ref{smear});
\begin{eqnarray}
{\sum_{l} \Tr\left\{M^{\da}(l)M(l)M^{\da}(l)M(l)\right\}} \; , & 
\displaystyle\sum_{l} \left(\Tr\left\{M^{\da}(l)M(l)\right\}\right)^2 \nonumber \\
\sum_{\left\langle l,l\prime\right\rangle} 
\Tr\left\{M(l)M(l^\prime)M^{\da}(l^\prime)M^{\da}(l)\right\} \; , & 
   \mbox{\em et cetera};
\label{zero}
\end{eqnarray}
$\langle l,l^\prime\rangle$ denotes a sum over neighboring
links. Note that gauge invariance 
\be
M(l) \to V^{\dagger}_{l-1,l} M(l) V_{l,l+1} \; , \; \; \;\;  V \in SU(N)
\eq
provides a strong constraint on the allowed
terms we can add at each order in the field $M$.
Contributions to the effective potential of the form
\be
  {N \nu \over \lambda}\, \Tr\left\{M(l)M(l+a)M^{\da}(l+a)M^{\da}(l)\right\} \; ,
\eq
where $\nu > 0$ and $l+a$ is the link consecutive to $l$
(parallel to $l$ and sharing one site), will be necessary 
in the small lattice spacing limit $\lambda \to 0$
in order to have any chance of 
producing the correct sign for the coefficient of the leading
rectangular $U$-plaquette correction: 
the  coefficient in Eqn.~(\ref{boxes}) now
becomes $\lambda(1-2 \nu)/g_{L}^{2} N$. Such terms have generally
been neglected in previous studies of dielectric lattice gauge theory.

We may instead 
integrate out the $U$-variables link-by-link  using the large-$N$
saddle-point result of Ref.~\cite{brezin}.
Thus, for some link $l$,
\begin{eqnarray}
\lefteqn{\int {\cal D}U\, \exp \left( - {N \over \lambda} \Tr \left\{
\left(M - U\right)\left(M^{\dagger} - U^{\dagger}\right)\right\} \right) = 
}\nonumber \\
& \exp \left( -N^2 x  - N {\displaystyle\sum_{i=1}^{N}} 
\left(\lambda \Lambda_i - 2
\sqrt{\Lambda_i + x}\right) - {1 \over 2} {\displaystyle\sum_{i,j =1}^{N}} \log
\left(\sqrt{\Lambda_i +x} + \sqrt{\Lambda_j +x}\right) -N^2
\left({3 \over 4}+ {1 \over \lambda} \right) \right) \label{bgint}
\end{eqnarray}
where $\Lambda_i$ are the eigenvalues of $MM^{\da}/\lambda^2$ and
\begin{eqnarray}
x = 0 \; \; & \mbox{if} & \; \; {1 \over N} 
\sum_{i=1}^{N} {1 \over \sqrt{\Lambda_i}} < 2
\nonumber \\
{1 \over N} 
\sum_{i=1}^{N} {1 \over \sqrt{x + \Lambda_i}} = 2 \; \; &\mbox{if}& \; \; 
{1 \over N}
\sum_{i=1}^{N} {1 \over \sqrt{\Lambda_i}} > 2 
\end{eqnarray}
Note that $\lambda$ and $\Lambda_i$ are $O(1)$ while the effective
action is $O(N^2)$ and $g^2_{L} N$ is, as usual, 
held finite in the large $N$ limit.
The asymptotics of the effective $M$-potential (\ref{bgint})
versus $\Tr\left\{M^{\da}M\right\}/N$
are easily found: 
for large eigenvalues $\Lambda_i$ we are in the 
$x=0$ regime and so (\ref{bgint}) becomes
\be
\exp \left( -{N \over \lambda} \Tr\left\{M^{\da} M\right\} \right) 
   \; , \; \; \; \;
 {1 \over N} \Tr\left\{M^{\da} M\right\} \to \infty
  \; ; \label{large}
\eq
for small $\Lambda_i$ we are in the $x \neq 0$ regime and  
may expand in powers of $\Lambda_i$
\be
x  =  {1 \over 4} - {1 \over N \lambda^2} \Tr\left\{M^{\da}M\right\} +
   O\left(M^4/\lambda^4\right)
\eq
to obtain
\be
\exp\left( \left( {-1 \over \lambda} + {1 \over \lambda^2} \right)N
 \Tr\left\{M^{\da}M\right\} + O\left(M^4/\lambda^4\right) \right) \; , 
\; \;\;\; {1 \over N} 
 \Tr\left\{M^{\da}M\right\}  \to 0 \; . \label{small}
\eq
As $\lambda$ is reduced through 1  the (classical) 
minimum shifts from $M=0$ to $\Tr\left\{M^{\da}M\right\} >0$.
We should not take too seriously the details of the
potential derived from (\ref{bgint})
beyond the Gaussian approximation, since the additional terms
(\ref{zero}) must then be taken into account. 


We see the qualitative picture
anticipated for the dielectric formulation emerging. At large enough lattice 
spacing (large enough $\lambda$),
the effective $U$-potential (\ref{boxes}), which contains
arbitrary Wilson loops with roughly equal weight, 
would be re-organised by
the dielectric variables into a  small-field expansion
about $M=0$. In practice one will have to truncate the 
small-field expansion and the form of the effective 
$M$-potential should be fixed by scaling analysis. It is difficult
to do this analytically because,  apart from the mass of the
$M$-field, there is no identifiably
small coupling at hand; $\lambda >1$ is not small, so we cannot
perform  improvement near the  fixed point, as is now
popular in the ELMC approach \cite{nied}. 
In fact, the ultimate justification for
expecting a truncation to work comes from light-cone quantisation,
as we argue in the next section.
%
%
(We will use an effective potential truncated to fourth order
in the fields (\ref{pot}).) 
We anticipate that
there should be an optimal  `window' for use of such a truncated
potential. A heavy $M$-field, when the minimum at $M=0$ is steep,
presumably implies that one is deep in the dielectric regime, where
the lattice spacing must be very large (analogous to the very
strong coupling regime of Wilson's theory);
in this case any truncation of higher order
terms in $V[M]$ may result in severe 
finite lattice spacing errors. 
On the other hand our Hamiltonian analysis will only be valid
for quantisation about the $M=0$ classical solution, 
so the mass squared  of the
$M$-field must be positive. This suggests we should take it
to be positive and small. Our results will bear this out.

Finally we note that one would arrive at basically the same
conclusions for an Abelian theory. However it is well-known that 
there is a phase transition between strong and weak coupling in this
case \cite{wilson,creutz}. The
dielectric formulation in the regime where $M$ has positive mass
squared would correspond to strong coupling, and is thus not useful
for studying weakly coupled QED.

\section{Light-cone Hamiltonian formulation}
\label{sec3}

\subsection{Transverse lattice}

A Hamiltonian limit of the dielectric lattice theory requires us to
take the lattice spacing to zero in at least one direction. 
We now specialise to $2+1$ dimensions, since it is for this case
that we will perform explicit calculations.
Since we also wish to make use of the fact that 
two components of the vector potential $A_{\mu}$ are
non-dynamical, we take
the continuum limit in the $x^0$ {\em and} $x^2$ directions, leaving
the transverse direction $x^1$ discrete.
This is means that only dynamical degrees of freedom will be involved
in the effective potential $V[M]$, a simplification not shared by
the Euclidean lattice or equal-time Hamiltonian quantisation of this
problem.
We now use different
smearing parameters $\lambda$ and $\tilde{\lambda}$ 
for the transverse $x^1$ and longitudinal $(x^0,x^2)$
directions, respectively; likewise we define transverse and longitudinal
lattice spacings $a$ and $\tilde{a}$.
For the longitudinal directions we take 
the limit $\tilde{\lambda},\tilde{a} \to 0$ so the measure becomes 
infinitely peaked on unitary matrices; this ensures that
\be
   \lim_{\tilde{\lambda},\tilde{a} \to 0} M(l) = U(l) = 
         1 + i\tilde{a}A_{\alpha} + \cdots
\label{limit}
\eq
for a link $l$ in the direction $\alpha = 0$ or $2$. 
Applying Eqn.~(\ref{limit}) to  $\Tr_{P}[M]$ 
one derives a transverse lattice action whose form
was first suggested in Ref.~\cite{bard}
\be
A  = \int dx^0 dx^2 \sum_{x_1}  \left(
  \Tr\left\{ D_{\alpha} M_{x^1}(D^{\alpha} 
M_{x^1})^{\da}\right\} - 
{a \over 4g^2} \Tr\left\{F_{\alpha\beta}F^{\alpha\beta}\right\}  
- V_{x^1}[M]\right)
\label{lag}
\eq
where
\be
D_{\alpha} M_{x^1}  =  \left(\partial_{\alpha} +i A_{\alpha} (x^1)\right)
        M_{x^1}  -  iM_{x^1} A_{\alpha}(x^1+a) 
\label{covdiv}
\eq
We have introduced the dimensionful continuum gauge coupling\footnote{
Note that this definition of $g^2$ differs from the standard
one by a factor of 2.} $g^2 =
a g^2_{L}/\tilde{a}^2$ in $2+1$ dimensions
and rescaled $M \to Mg_L$.
$M_{x^1}$ lies on the link between $x^1$ and $x^1 + a$ while
$A_{\alpha}(x^1)$
is associated with the site $x^1$. $V_{x^1}[M]$ is a purely transverse
gauge invariant effective potential, 
which is assumed analytic about its minimum at $M=0$. 
For $2+1$ dimensions this means it contains only 
(products of) zero-area Wilson loops such as
those displayed in (\ref{zero}) with one point pinned at $x^1$. 

A feature of light-cone formalism, which is not present in the 
equal-time approach, is that for free fields of mass squared $\mu^2 > 0$
the momentum $k^+ = (k^0 + k^2)/\sqrt{2}$ is positive definite for finite
energy since
\be
k^- = {\mu^2 + (k^1)^2 \over 2k^+}
\eq
This, together with momentum conservation, implies that even the 
quantum vacuum is free of particles and that a large number
of particles necessarily carry a large light-cone energy. It is for
this reason that one believes a calculation of the low mass spectrum
using the light-cone Hamiltonian is feasible when effects due to
large numbers of particles have been truncated. Higher powers of the
field $M$ in the effective potential couple precisely to those
components of the wavefunction with a large number of $M$-quanta. Of
course, in the \LCTL{} construction, the $M$-quanta are also link
variables, so energy considerations are intimately
tied in with questions of spacetime symmetry restoration.
Note that the above arguments break down if applied to the gluon fields
$A_1$ directly since they are massless and can be produced
copiously for $k^1 =0$. On the other hand, if one gives the gluons
$A_1$ a fake mass this breaks gauge invariance leading to
copious production of non-local terms in the effective potential.

\subsection{Gauge symmetry}

Having formulated a transverse lattice action (\ref{lag}) we now 
introduce light-cone co-ordinates $x^{\pm} = x_{\mp} = (x^0 \pm x^2)/\sqrt{2}$
and quantise by
treating $x^+$ as canonical time, following Ref.~\cite{bard}. 
The theory has a conserved current
%
%
\be
%
   J^{+}({\bf x})  =  i \left[
        M_{x^1} 
        \left(D_- M_{x^1} \right)^{\da}  - 
           \left(D_-  M_{x^1} \right) M_{x^1}^{\da}+
           M_{x^1-a}^{\da} D_- M_{x^1-a}  - 
          \left(D_-  M_{x^1-a}\right)^{\da}
           M_{x^1-a} \right] \label{J}
\eq
at each transverse lattice site $x^1$ which plays a special role.
If we pick the light-cone gauge\footnote{We set the associated
zero mode $\int dx^- \, A_-$ to zero as a dynamical approximation.} $ A_- =0$ 
the non-propagating field $A_+$ 
satisfies a simple constraint equation at each transverse site
\be
\left(\partial_{-}\right)^{2} A_+(x^1) = 
{g^2\over a} J^{+}_{x^1} \; . \label{constr}
\eq
Solving this constraint leaves an action in terms of 
the dynamical fields  $M_{x^1}$
\be
A =  \int dx^+ dx^- \sum_{x^1} \Tr \left\{
\partial_{\alpha} M_{x^1} \partial^{\alpha} M_{x^1}^{\dagger} 
+ {g^2 \over 2a} 
J^{+}_{x^1} \frac{1}{\left(\partial_-\right)^2} J^{+}_{x^1} \right\}
- V_{x^1}[M] \; ,  \label{action}
\eq
the pole in the $A_+$ propagator given by $1/(\partial_{-})^2$
is regulated by a principle value type prescription \cite{bard,thooft}. 
It is straightforward to derive
light-cone momentum and energy
\begin{eqnarray}
  P^+ & = & 2 \int dx^- \sum_{x^1} \Tr  \left\{ \partial_- M_{x^1} 
\partial_- M_{x^1}^{\da} \right\} \label{mom} \\
 P^-  & = & \int dx^- \sum_{x^1} V_{x^1}[M] - {g^2 \over 2a} \Tr \left\{ 
      J^{+}_{x^1} \frac{1}{\partial_{-}^{2}} J^{+}_{x^1}
           \right\} \label{energy}
\end{eqnarray}
There remains residual gauge symmetry under $x^-$-independent
transformations at each site $x^1$
\be
M_{x^1} \to V^{\dagger}_{x^1+a} M_{x^1} V_{x^1} \; , \;\;\;\; V \in SU(N)
 \label{sym}
\eq
Evidently the light-cone energy (\ref{energy})
will not be finite unless the associated
charge vanishes at  each site $x^1$
\be
\int_{-\infty}^{+\infty} dx^- J^{+}_{x^1} = 0 \label{charge}
\eq
This requirement also follows from the zero momentum
limit of the constraint equation (\ref{constr}) and will force one to consider
only singlet states under the transformations (\ref{sym}). This
condition is analogous to the colour neutrality imposed by Kogut and
Susskind \cite{kogsus1} as a consequence of Gauss' law, but here it
follows from dynamical considerations alone.

\subsection{The Hilbert space}

Expanding in harmonic modes at $x^+ =0$ about $M=0$, 
it is convenient to work in
longitudinal momentum space and transverse position space
\be
 M_{x^1}(x^-)  =  
\frac{1}{\sqrt{4 \pi }} \int_{0}^{\infty} {dk \over {\sqrt k}}
    \left( a_{-1}(k,x^1) e^{ -i k x^-}  +  \left(a_{+1}(k,x^1)\right)^{\da} 
e^{ i k x^-} \right)
\eq
where the modes satisfy equal-$x^+$ commutators
\begin{eqnarray}
\left[a_{\lambda,ij}(k,x^1), (a_{\rho,kl}(\tilde{k},\tilde{x}^1))^{\da}\right] 
& = & \delta_{ik} \delta_{jl} \delta_{\lambda \rho} \delta_{x^1\tilde{x}^1}
\delta(k-\tilde{k}) \\
\left[a_{\lambda,ij}(k,x^1),a_{\rho,kl}(\tilde{k}, \tilde{x}^1)\right] & = & 0
\end{eqnarray}
with colour indices $i,j \in \{1,\ldots,N\}$, 
orientation indices $\lambda, \rho \in \{-1,+1\}$;
$\left(a_{\lambda,ij}\right)^{\da} = (a^{\da}_{\lambda})_{ji}$ and for
clarity we omit the $+$ superscript on $k$. The Fock space operator
$a_{\pm 1}^{\da}(k,x^1)$ creates a mode on the link between sites $x^1$ and
$x^1 + a$ with longitudinal momentum
$k$ and orientation $\pm 1$. 
In the Fock space, only combinations singlet under
(\ref{sym}) are annihilated by the charge (\ref{charge}).
This gives a Hilbert space at fixed $x^+$ formed from 
all possible closed Wilson loops of link
modes  $a_{\pm}$ on the transverse lattice. 
Thus, a typical $p$-link loop will be something like
\be
\Tr  \left\{ a_{+1}^{\da}(k_1,x^1) a_{-1}^{\da}(k_2,x^1)
a_{-1}^{\da}(k_3,x^1-a)
\cdots a_{+1}^{\da}(k_p,x^1-a)\right\} \,  |0\rangle \label{typ}
\eq
The number of $+1$'s equals the number of $-1$'s for a closed loop on
the transverse lattice, while the longitudinal momenta $k_m$ are
unconstrained except for $\sum_{m=1}^{p} k_m = P^+$ and $k_m \ge 0$.
By an abuse of language, we will often refer to $p$ as the number of 
particles --- the link variables are reminiscent of constituents ---
even though the relevant field variables are non-locally defined.
At large $N$ we need only study the dynamics of
single connected Wilson loops in the Hilbert space --- there are no
terms with more than one trace ---  since the 
loop-loop coupling constant is of order $1/N$.
These loops may be thought
of as `bare' glueballs, and the problem is to find the linear
combinations that are on mass shell.
Neglecting $k_m=0$, 
which is consistent
with expanding about the $M=0$ solution of the dielectric regime,
the Fock vacuum is an eigenstate of the full
light-cone Hamiltonian 
$P^- \left|0\right\rangle = P^+ \left|0\right\rangle =0$.

\subsection{Eguchi-Kawai reduction}

A further simplification occurring at large $N$ is Eguchi-Kawai reduction
\cite{ek}. This means that in the Lorentz frame $P^1=0$ the theory 
is isomorphic to one compactified  on
a one-link transverse lattice 
with periodic boundary conditions, where $P^\alpha$ acts on
a basis of zero winding number Wilson loops \cite{dalley1}, 
{\em id est} we can 
simply drop the argument $x^1$ or $l$ from  $M$ in all of the previous
expressions. Effectively one is now dealing with a $1+1$-dimensional 
gauge theory coupled to a complex scalar field in the adjoint 
representation (with self-interactions).   
Mathematically isomorphic ``collinear QCD
models'' have recently been studied \cite{AD}, 
the  equivalence following after identifying winding number (orientation) 
here with  helicity there.

The Eguchi-Kawai reduced states corresponding
to $P^1 =0$ and fixed $P^+$ can now be written as colour
singlet Fock basis states  on one link. A general state expanded in terms 
the orthonormal basis is then
(summation on repeated indices implied)
\begin{eqnarray}
|\Psi\rangle = \sum_{\begin{array}{@{}c@{}}
    \scriptstyle p=|n|,|n|+2,\ldots \\
    \scriptstyle p>0\end{array}} 
\int_{0}^{P^+} {dk_1 \cdots dk_p\over N^{p/2}} \;
 \delta \! \left(P^+ - \textstyle\sum_{m=1}^p k_m\right) \,
f^{\lambda \rho \ldots \sigma}(k_1,\ldots,k_p) \nonumber \\
\cdot \Tr \left\{ a_{\lambda}^{\da}(k_1) a_{\rho}^{\da}(k_2)\cdots
a_{\sigma}^{\da}(k_p)\right\} \,  |0\rangle \label{wf}
\end{eqnarray}
where we set the winding number $n=  \lambda + \rho + \cdots + \sigma$ 
equal to zero. 
When the wavefunctions $f$ 
are chosen appropriately, the states $ |\Psi\rangle$ will 
describe an infinite  discrete set of stable physical glueballs satisfying
the mass shell condition. 

Using the mode expansion of the operators $P^\alpha$ in
Fock space, one may explicitly verify 
that the isomorphism under reduction holds. Some details of this 
calculation and its validity are given in Appendix~\ref{appendixa}.
The reduction rests on expanding about the $M=0$ classical solution. 
We have already argued that, at sufficiently small lattice
spacing, the classical minimum is no longer at $M=0$, 
and our quantisation is no longer valid. Note that this is not a statement
about equivalence under reduction breaking down, but  about our entire
method, whether applied to the reduced or unreduced theory.
Wilson's lattice  gauge theory 
in the Lagrangian formulation is valid for any correlation length, but
Eguchi-Kawai reduction breaks down at small lattice spacing
\cite{twist} and
various quenching and twisting prescriptions have been given in order to
approach the critical surface in coupling constant space
with a reduced theory. The
dielectric formulation with expansion about $M=0$  does not need
quenching/twisting, but we pay for this by having to work with an effective
theory at rather  large lattice spacing on the scaling trajectory.

\subsection{Symmetries} 

The theory possesses several discrete symmetries. Charge conjugation
induces the symmetry 
\be
{\cal C}: \, a_{+1,ij}^{\da} \leftrightarrow a_{-1,ji}^{\da} \; .
\eq
There are two orthogonal reflection symmetries ${\cal
P}_1$ and ${\cal P}_2$ either of which may be used as `parity'.
If ${\cal P}_1:  x^1 \to -x^1$,  we have 
\be
{\cal P}_1:  \, a_{+1,ij}^{\da} \leftrightarrow a_{-1,ij}^{\da} \; .
\eq
${\cal P}_2$: $x^2 \to -x^2$, is more subtle in light-cone
quantisation since it does not preserve the quantisation surface. 
Its explicit operation is known for sure only for a set of $p$ free
particles \cite{horn}, which we call 
\be
{\cal P}_{2f}: \, {k_m \over P^+} 
\to \left( k_m \sum_{m'=1}^{p} {1 \over k_{m'}}
\right)^{-1} \label{pfree} \; .
\eq
The latter  $P_{2f}$ is
nevertheless useful since it is often an approximate quantum number
and its expectation value can be used to estimate ${\cal P}_2$ 
\cite{AD,bvds}.
Given ${\cal P}_2$ and ${\cal P}_1$ we can determine whether spin
${\cal J}$ is even or odd using the relation 
$(-1)^{{\cal J}} = {\cal P}_1 {\cal P}_2$.
If rotational symmetry has been restored in the theory, 
states of spin ${\cal J} \neq 0$ should 
form degenerate ${\cal P}_1$ doublets 
$|+{\cal J}\rangle \pm |-{\cal J}\rangle$~\cite{teper}.
We use ``spectroscopic notation'' 
$|{\cal J}|^{{\cal P}_1 {\cal C}}$ to classify states.

\subsection{The effective potential}

The basis written in (\ref{wf}) is already diagonal in $P^+$, and $P^+$
commutes with $P^-$.
It remains to
find the coefficient functions $f$, cyclically symmetric in their
arguments,
 which diagonalise  the Hamiltonian $P^-$ and hence
the Lorentz invariant $\left(\mbox{mass}\right)^2$ 
operator $2 P^+ P^-$ with eigenvalue $M^2$.
Before proceeding we must choose the explicit
form of the transverse effective potential.
In this work, we will include in $V[M]$ {\em all} Wilson loops and products
of Wilson loops up to fourth order in 
link fields $M$. After  Eguchi-Kawai  reduction this limits one
to the following form
\begin{eqnarray}
  V[M] & = & \mu^2  \Tr \left\{MM^{\da}\right\} + {\lambda_1 \over a N}
\Tr \left\{MM^{\da}MM^{\da} \right\} \nonumber  \\
&&+ {\lambda_2 \over a N}  \Tr \left\{MM^{\da}M^{\da}M\right\}
+ {\lambda_3 \over a N^2} \Tr \left\{M^{\da}M\right\} 
\Tr \left\{M^{\da}M\right\} 
\label{pot}
\end{eqnarray}
Note that the last term above, which might appear suppressed at
large $N$, is in fact non-zero only for 2-link Fock states ($p=2$ in 
Eqn.~(\ref{wf})). It is a 
term which `pinches' the worldsheet swept out in light-cone time
$x^+$  by the glueball
flux ring given by the superposition of Wilson loops (\ref{wf}). 
Near a critical surface in coupling constant space
it is naively irrelevant, but at 
finite lattice spacing it turns out to be quite significant.
In principle one should repeat calculations with higher orders of the
field allowed in the action, to determine the magnitude of any changes. 
We will not be able to
do this with our modest computing resources --- there are many more allowed
terms at 6th order --- but will look at other quantities, such as
expectation values of higher order Wilson loops, in order to assess
effects of this truncation. (One can also of course compare with results
obtained using only the quadratic term in (\ref{pot})). Our goal is to find out
if there is a trajectory in the parameter space of (\ref{pot}) along which
physical quantities scale {\em and} agree with the ELMC simulations 
\cite{teper} with reasonable $\chi^2$.

The renormalisation of the quantum theory at fixed lattice spacing 
follows that of a 2D gauge
theory with adjoint 
matter \cite{AD},
%
%
involving only logarithmically divergent
self-energy subtractions on the propagator and a principle value
prescription for the light-cone Coulomb singularity. 
Explicitly one finds at large $N$
\begin{eqnarray}
P^+ & = & \int_{0}^{\infty} dk \, k \, 
{\rm Tr}\left\{ {a^{\da}_{\lambda}(k)} a_{\lambda}(k) \right\} \\
P^- & = & {\mu^2 \over2 } \int_{0}^{\infty} {dk \over k}
 {\rm Tr}\left\{ {a^{\da}_{\lambda}(k)} a_{\lambda}(k) \right\}
\nonumber \\  
& & + {\rm P}\int_{0}^{\infty} 
\frac{dk_1 \, dk_2 \, dk_3 \, dk_4}{8 \pi a \sqrt{k_1 k_2 k_3 k_4}}  
[ \delta(k_1 + k_2 - k_3 - k_4) {\cal H}_{2 \rightarrow 2}
\nonumber \\ &&
+ \delta(k_1 + k_2 + k_3- k_4)\left( {\cal H}_{1 \to 3}
+ {\cal H}_{3 \to 1}\right)]\label{minus2}
\end{eqnarray}
where
\begin{eqnarray}
{\cal H}_{2 \to 2} & = & \left(\frac{\lambda_1}{N} - \frac{g^2
   \left(k_2-k_1\right) \left(k_4-k_3\right)}
   {\left(k_1+k_2\right)^2}\right)
   \Tr\left\{a^{\da}_{\lambda}(k_1) a^{\da}_{-\lambda}(k_2)
   a_{\rho}(k_3) a_{-\rho}(k_4) \right\} \nonumber \\
 & & +\left(\frac{\lambda_1}{N} - \frac{g^2
   \left(k_2+k_3\right) \left(k_1+k_4\right)}
   {\left(k_2-k_3\right)^2}\right)
   \Tr\left\{a^{\da}_{\lambda}(k_1) a^{\da}_{\rho}(k_2)
   a_{\rho}(k_3) a_{\lambda}(k_4) \right\} \nonumber \\
 & & + \frac{\left(\lambda_2-\lambda_1\right)}{N}
   \Tr\left\{a^{\da}_{\lambda}(k_1) a^{\da}_{\rho}(k_2)
   a_{\lambda}(k_3) a_{\rho}(k_4) \right\} \nonumber \\
& & + {\lambda_3 \over 2N^2 }
\Tr\left\{a^{\da}_{\lambda}(k_1)a^{\da}_{-\lambda}(k_2)\right\}
\Tr\left\{a_{\rho}(k_3)a_{-\rho}(k_4)\right\} \label{coul} \\
{\cal H}_{1 \to 3} & = & \left(\frac{\lambda_1}{N} - \frac{g^2
   \left(k_2-k_1\right) \left(k_4+k_3\right)}
   {\left(k_1+k_2\right)^2}\right)
   \Tr\left\{a^{\da}_{\lambda}(k_1) a^{\da}_{-\lambda}(k_2)
   a^{\da}_{\rho}(k_3) a_{\rho}(k_4) \right\} \nonumber \\
 & & +\left(\frac{\lambda_1}{N} - \frac{g^2
   \left(k_2-k_3\right) \left(k_1+k_4\right)}
   {\left(k_2+k_3\right)^2}\right)
   \Tr\left\{a^{\da}_{\lambda}(k_1) a^{\da}_{\rho}(k_2)
   a^{\da}_{-\rho}(k_3) a_{\lambda}(k_4) \right\} \nonumber \\
 & & + \frac{\left(\lambda_2-\lambda_1\right)}{N}
   \Tr\left\{a^{\da}_{\lambda}(k_1) a^{\da}_{\rho}(k_2)
   a^{\da}_{-\lambda}(k_3) a_{\rho}(k_4) \right\} \label{create} \\
{\cal H}_{3 \to 1} &=& {\cal H}_{1 \to 3}^{\da} \; .
\end{eqnarray}
In the $N=\infty$ limit we do not distinguish
$U(N)$ from $SU(N)$ and the gauge coupling $g^2 N$ is finite. 
The light-cone Schr{\"o}dinger equation defined by taking the
matrix elements $\left\langle \mbox{\rm Fock}| 2P^+P^- |\Psi\right\rangle = 
M^2 \left\langle \mbox{\rm Fock} |\Psi \right\rangle$ 
allows one to write down an infinite set of
successively coupled integral equations for the corresponding 
Fock components $f$. We do not bother to list them here since
they are rather complicated, but refer the reader to Ref.~\cite{AD} for
equivalent expressions.
To diagonalise $P^-$ we will employ a combination of
variational ans\"{a}tze  for the $f$'s and numerical solution through
discretisation the momenta $k$ and truncation of the number of
particles $p$. 
%

\section{The glueball spectrum}
\label{sec4}
\subsection{Analytic wavefunctions}

The behavior of the wavefunctions $f$ in Eqn.~(\ref{wf})
when any one of the arguments vanishes~\cite{bard} is 
\be
 \lim_{k_i \to 0}
 f_{\lambda \rho \ldots \sigma}(k_1, k_2, \ldots , k_p) 
    \propto k_{i}^{\beta} \;\;\;\; \mbox{where} \;\;\; \;  
    2 \beta \tan{(\pi \beta)} = {a \mu^2 \over g^2 N} 
     \label{vanish}
\eq
and $0 < \beta < 1/2$.
Specifying also the number of nodes of $f$ as
a function of momenta,
one can make a sensible ansatz. For $\lambda_1$ and $\lambda_2$ small
there is very little mixing between Fock states of  different number
of link modes $p$ \cite{AD}.
In this case a mass eigenstate $|\Psi\rangle$ has predominantly a 
fixed $p$, the mass increasing with $p$.
For a given $p$, the energy also tends to increase with the number of
nodes in the wavefunction $f$ due to the $J\left(\partial_-\right)^{-2}J$ 
term in (\ref{energy}),
which is in fact a positive contribution. 
Thus one naively expects the lowest
two eigenstates to be approximately
\be
\int_{0}^{P^+} dk \, f_{+1,-1}(k,P^{+} - k) \, \Tr 
\left\{ a_{+1}^{\da}(k) a_{-1}^{\da}(P^+-k) \right\} |0\rangle
\eq
with the lowest state having a symmetric wavefunction $f_{+1,-1}(k,P^+-k)$,
corresponding to $0^{++}$, 
and first excited state having $f_{+1,-1}$ antisymmetric with one node, 
corresponding to $0^{--}$. 
The next highest states are either a
4-link state with positive symmetric wavefunctions $f_{+1,+1,-1,-1}$
and  $f_{+1,-1,+1,-1}$ or a symmetric
2-link state with $f_{+1,-1}$ having two nodes. In the glueball spectrum
we  identify the latter states as $0^{++}_{*}$ and $2^{++}$, respectively,
although actual eigenstates are a mixture of these.
The above naive considerations correctly predict the ordering of light
glueballs in the spectrum, but differ in the details of the wavefunctions
we find numerically later.

\label{transverseonly}

In light of the endpoint analysis (\ref{vanish}),
it is useful to consider wavefunctions $f$ 
in Eqn.~(\ref{wf}) with the simplified longitudinal 
form $k_i^\beta$ and arbitrary transverse structure:
a ``transverse only'' model.  
Thus we construct a basis of unit norm states of the form:
\begin{eqnarray}
 \left| \sigma_1 ,\ldots , \sigma_p \right\rangle =
 C_{\sigma_1,\ldots , \sigma_p}
   \sqrt{\frac{\Gamma(2 \beta p+p)}{N^p{\Gamma(2 \beta+1)}^p}}
  \int dk_1 \cdots dk_p \; \Tr
   \left\{ a^{\dag}_{\sigma_1}(k_1) \cdots  a^{\dag}_{\sigma_p}(k_p)
    \right\} \left|0\right\rangle \nonumber \\
  \cdot \delta\left(p^+-\sum\nolimits_i k_i\right) 
         \left(k_1 \cdots k_p\right)^\beta 
       \label{trbasis}
\end{eqnarray}
where $\sigma_i \in \{-1,+1\}$ and we set the winding number
 $n=\sum_i \sigma_i$ to zero.
One can characterise these states with the 
``folding parameter'' $\gamma$:
\begin{equation}
  p \gamma = \mbox{The number of $i$ such that $\sigma_i = \sigma_{i+1}$.}
\end{equation}
Thus,
\begin{equation}
  p (1-\gamma) = \mbox{The number of $i$ such that $\sigma_i \ne \sigma_{i+1}$.}
\end{equation}
(Here, one should think of a 1-dimensional polymer model.)
Roughly speaking, the transverse size of a 
typical state is $a \left(p/2\right)^\gamma$.
Using $\gamma$, we can write down an exact 
expression for the expectation value of the Lorentz
invariant $(\mbox{mass})^2$ operator:
\begin{eqnarray}
\lefteqn{  \left\langle \sigma_1 ,\ldots , \sigma_p \right| P_\mu P^\mu
 \left| \sigma_1 ,\ldots , \sigma_p \right\rangle}\nonumber \\
  &=&  p \left(p(1+2 \beta)-1\right) \cdot
 \nonumber \\  & & \left(\frac{\mu^2}{2 \beta} 
    + \frac{
       {\Gamma(\beta+1/2)}^4 \Gamma(1+4 \beta)
    }{
      4 \pi a {\Gamma(1+2 \beta)}^4
     } \left[g^2 N(1+4 \beta)+2 (1-\gamma) \lambda_1 + \gamma \lambda_2 \right]
    \right) \; , \; \; p>2 \label{tro}\\
 &=& \frac{p^2\pi}{4 a}\left(g^2 N+2 (1-\gamma) \lambda_1
             +\gamma \lambda_2\right)
    \nonumber \\
 &&   +\frac{\beta \pi p^2}{2 a} \left(g^2 N \left(5-4 \ln 2\right)
 -\left(2 \lambda_1 (1-\gamma)+\lambda_2 \gamma\right)
         \left(4 \ln 2 -1\right)
    \right)\nonumber \\
 &&   +O\left(p\right)+O\left(\beta^2\right) \; .
\end{eqnarray}
For the two particle state $\left|+1,-1\right\rangle$
the expectation value of the invariant $(\mbox{mass})^2$
operator is,
\begin{eqnarray}
\left\langle +1,-1 \right| P_\mu P^\mu\left|+1,-1\right\rangle
 &=& \frac{\pi}{2 a}\left(g^2 N+2  \lambda_1 + \lambda_2 + \lambda_3\right)
    \nonumber \\
 &&   +\frac{2 \beta \pi}{a} \left(g^2 N \left(3-2 \ln 2\right)
 -\left(2 \lambda_1+\lambda_2+ \lambda_3\right)
         \left(2 \ln 2 -1\right)
    \right)\nonumber \\
 &&   +O\left(\beta^2\right) \; .
\end{eqnarray}
If one demands a positive-definite spectrum, these expressions
place variational bounds on the allowed region in parameter space.
%
%
First we must have 
positive $\mu^2$ else our quantisation procedure is no longer valid
and the spectrum becomes unbounded from below.
For the $p=2$ case the 
expectation value of the invariant $(\mbox{mass})^2$
becomes tachyonic when
\begin{equation}
   2 \lambda_1 + \lambda_2 + \lambda_3 \le
    -g^2 N \left(1+8 \beta\right) +O\left(\beta^2\right) \; ,
\end{equation}
while from (\ref{tro}) we see that many particle states become tachyonic at
\begin{equation}
   2 (1-\gamma) \lambda_1 + \gamma \lambda_2 \le
    -g^2 N \left(1+8 \beta\right) +O\left(\beta^2\right) \; . \label{bounds}
\end{equation}

\subsection{Numerical results}

In our numerical solutions we restrict the number of link fields $p$ 
in our basis states (\ref{wf}) and discretise momenta
by demanding antiperiodicity of the fields in $x^- \to x^-+L$. This is
a high light-cone energy cut-off since, unlike in the transverse
direction,
light-cone momentum goes like the inverse of energy.
(we use the method of \cite{AD} to handle the IR divergence).
For fixed integer valued cut-off $K= L P^+/(2\pi)$ momenta
are labeled by odd half integers $\kappa_m = K k_m/P^+$, 
$K = \sum_m \kappa_m$.  On the computer, we generate
a basis of states and calculate matrix elements of the Hamiltonian
using the hybrid wavefunction/DLCQ matrix elements discussed
in Appendix~\ref{appendixc}.  The nonzero matrix elements are
then stored in a file (the matrix is, in fact, 
quite sparse).  To diagonalise the Hamiltonian $P^-$, we employ
a standard Lanczos algorithm. 

\begin{figure}
\centering
\BoxedEPSF{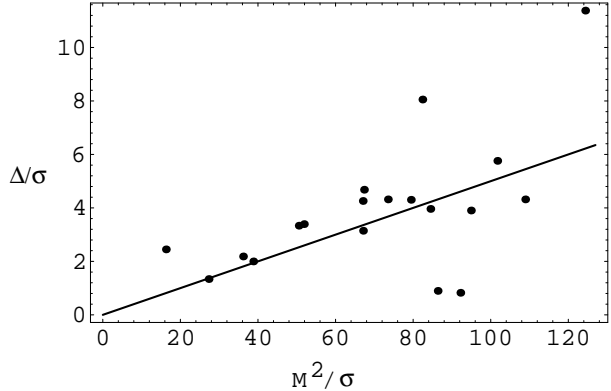 scaled 667}
\caption{
Difference between the unextrapolated and extrapolated 
eigenvalues
$\Delta = \left|M^2_{K=14,p \le 6}-M^2_{\mbox{\footnotesize 
extrapolated}}\right|$ vs $M^2$ 
for the spectrum in Fig.~\protect\ref{fig4}.
Also shown is the function $0.05 M^2$ which is
the assumed size for our error bars.
\label{fig44}}
\end{figure}

In order to fix the coupling constants in the effective 
potential, we perform a least $\chi^2$ fit to Teper's ELMC 
large $N$ extrapolated spectrum.\footnote{For the ELMC data, 
we assumed rotational symmetry and averaged over parity doubles.}  
As we shall later show,
the mass in units of the coupling $m^2= \mu^2 a/(g^2 N)$
is a measure of the lattice spacing while the other terms
in our effective potential $\lambda_1$, $\lambda_2$,
and $\lambda_3$ must be found from the fitting procedure.
%
We will also determine $g^2 N/a$ based on a fit, which we check
self-consistently with measurements of the string tension.

In order to minimise errors associated with our
truncation in $p$ and $K$, we take the spectra for 
\be
    \left(K,\mbox{$p$ truncation}\right) \in 
     \left\{ (10,4), (10,6),(10,8),(11,6),
             (12,6), (13,6),(14,6)\right\}
\eq
and extrapolate to the continuum, $K,p \to \infty$, 
using a least squares fit to the function
\be
       c_0 + \frac{c_1}{K} + c_2 \, e^{c \, p}
         \label{ccc}
\eq
where $c$ is given by our estimate in Appendix~\ref{appendixb} and 
the $1/K$ term is consistent with the leading  
finite $K$ error as discussed in Appendix~\ref{appendixc}.
Since we cannot measure $|{\cal J}|$ directly,
we only classified states according to ${\cal P}_1$ and  ${\cal C}$
during the fitting process.
In practice we found that the fitting function (\ref{ccc}) worked quite
well for all but the highest 3 or 4 states where the 
exponential convergence in $p$-truncation was not yet 
evident.\footnote{However, looking at the $K=10$, $p \le 8$ sector, even 
these highest states had less than 15\% 8-particle content.}
We estimate our one-sigma errors from finite $K$ and $p$-truncation 
to be roughly $0.05 M^2$; see Fig.~\ref{fig44}.

%
%
\begin{table}
\begin{center}
\begin{tabular}{c|ccc|cc}
$\displaystyle m^2=\frac{\mu^2 a}{g^2 N}$ & 
$\displaystyle \frac{\lambda_1}{g^2 N}$ & 
$\displaystyle \frac{\lambda_2}{g^2 N}$ &
$\displaystyle \frac{g^2 N}{a \sigma}$ & 
$\displaystyle \chi^2$ & $\displaystyle c$ \\[10pt]
\hline
 0.      & -0.00565 & -0.305 & 4.38 & 30.  & -1.43  \\
 0.00394 & -0.035   & -0.366 & 4.11 & 25.9 & -1.29  \\
 0.0158  & -0.0677  & -0.433 & 3.84 & 25.3 & -1.2   \\
 0.036   & -0.123   & -0.496 & 3.65 & 23.1 & -1.1   \\
 0.065   & -0.202   & -0.55  & 3.47 & 23.5 & -1.01  \\
 0.104   & -0.297   & -0.602 & 3.28 & 23.6 & -0.94  \\
 0.153   & -0.427   & -0.626 & 3.1  & 24.9 & -0.884 \\
 0.214   & -0.58    & -0.658 & 2.97 & 25.7 & -0.832 \\
 0.291   & -0.778   & -0.657 & 2.84 & 28.6 & -0.792 \\
 0.384   & -1.05    & -0.508 & 2.67 & 35.2 & -0.802 \\
 0.5     &  ---     &  ---   & ---  & ---  & ---
\end{tabular}
\end{center}
\caption{Scaling trajectory for the extrapolated spectrum.
Here we tabulate couplings vs $m^2$ for a best $\chi^2$ fit
to the ELMC spectrum.
We find that $\lambda_3/(g^2 N)$ is always quite
large ($>100$).
We could not find a satisfactory minimum
for the $m^2=0.5$ datum.
\label{scaling}}
\end{table}

\begin{figure}
\begin{tabular}{@{}c@{}c@{}}
\BoxedEPSF{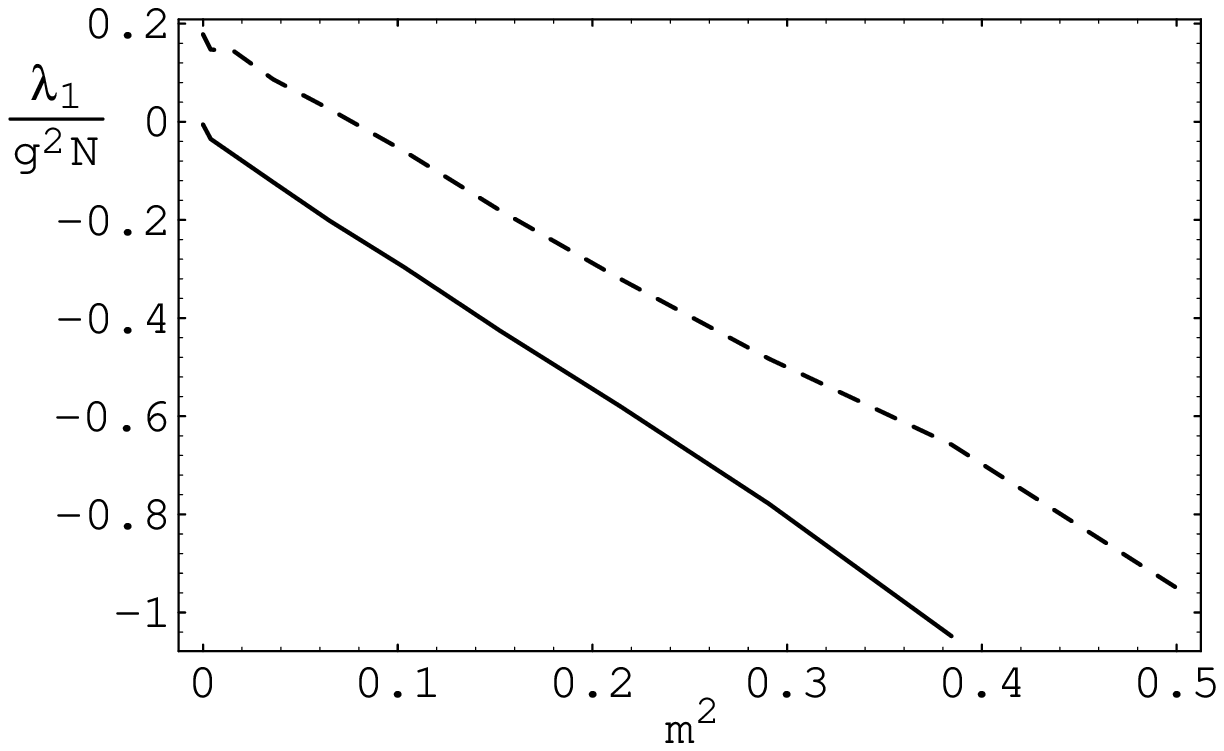 scaled 590} &
\BoxedEPSF{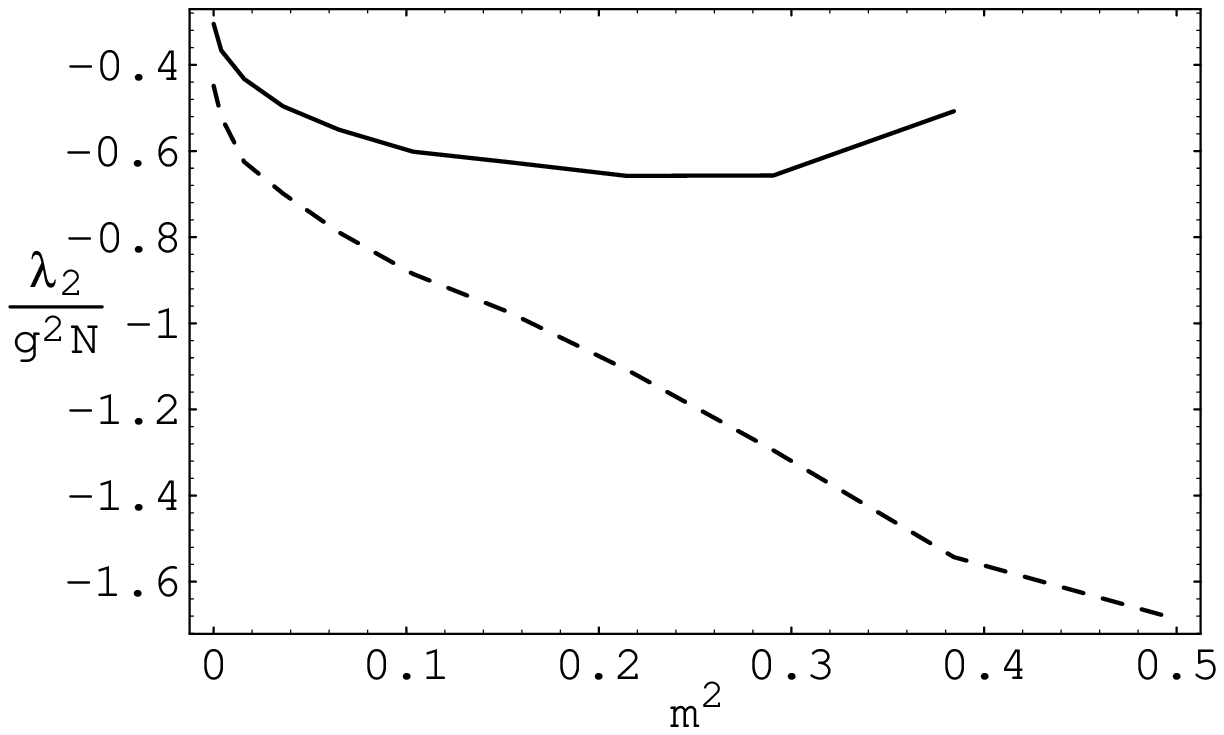 scaled 590} \\
  (a)  & (b)
\end{tabular}
\caption{Scaling trajectory in parameter space.
We plot (a) $\lambda_1/(g^2 N)$ and (b) $\lambda_2/(g^2 N)$
vs mass $m^2$.
The solid line is the data in Table~\protect\ref{scaling};
the dashed line is from a second trajectory discussed
in Section~\protect\ref{question}.
 \label{fig12}}
\end{figure}

As we fit the various couplings as a function of $m^2$, we
find a narrow strip in parameter space 
where we obtain
good agreement with the ELMC spectrum; see Table~\ref{scaling}
and Fig.~\ref{fig12}.
As we shall see, moving along this strip corresponds
to changing the lattice spacing $a$. The strip, where $\chi^2$ has a
local minimum,
disappears when  $m^2$  is sufficiently large, 
indicating that for large enough lattice spacing
our truncation of the effective potential is no longer a good approximation.

%
%
%
%
\begin{figure}
\centering
\BoxedEPSF{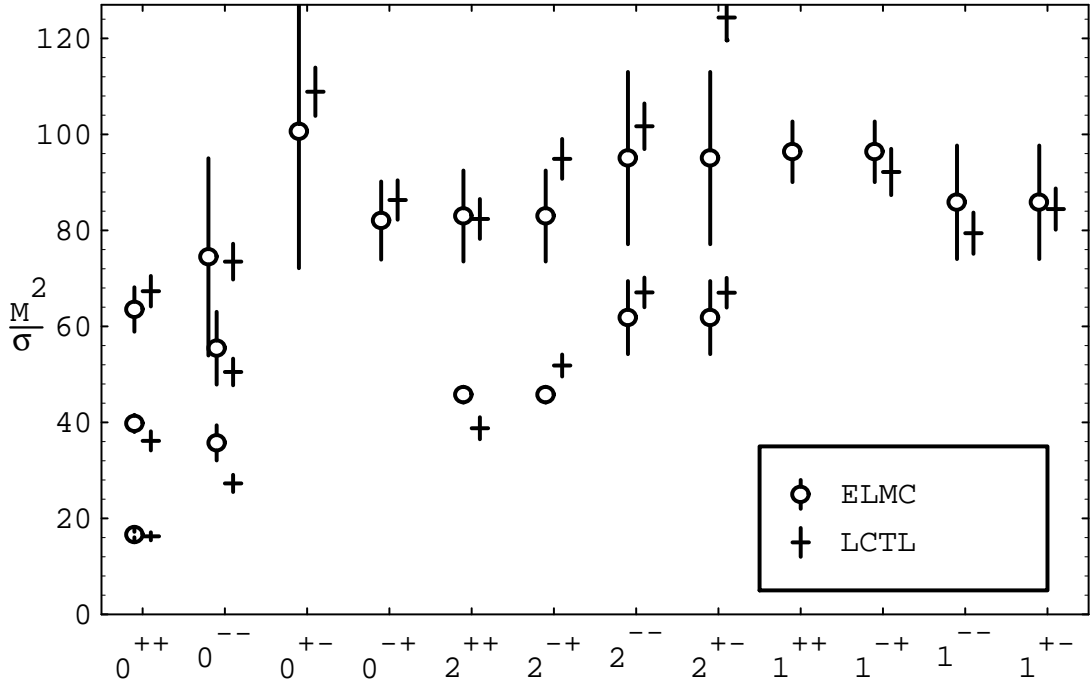 scaled 1000}
\caption{
A fit of our \LCTL{} extrapolated results against
Teper's ELMC large $N$ extrapolated spectrum 
\protect\cite{teper}.
The $M^2$ eigenvalues are shown in units of (Teper's) string tension
for various ${\left|{\cal J}\right|}^{{\cal P}_1 {\cal C}}$.
The parameters are from the $m^2=0.065$ row of 
Table~\protect\ref{scaling}.
One finds similar spectra along the entire scaling trajectory in
coupling constant space.
\label{fig4}}
\end{figure}

In Fig.~\ref{fig4}, we have plotted a typical spectrum along
the scaling trajectory together with the  ELMC 
results.  For graphing purposes, we assigned $\left|{\cal J}\right|$ 
to our spectrum based on a best fit to Teper's results.
As a partial check of this assignment, we calculated the 
expectation value of ${\cal P}_{2 f}$, Eqn.~(\ref{pfree}), for our 
spectrum ($K=10$, $p \le 6$ truncation).  
The subsequent determination of $(-1)^{\cal J}$ agreed with the spectrum
fit for most of the low-lying glueballs.
However, there was disagreement
in the assignment of $(-1)^{\cal J}$ between the lowest $0^{-+}$, 
first excited $2^{-+}$, and lowest $1^{-+}$ states in Fig.~\ref{fig4}. 
However, since these three levels are almost degenerate, this discrepancy
is not important. Of more importance is the $\left|{\cal J}\right|^{+-}$ 
sector where our measurements were ambiguous 
(with $\left\langle\Psi\right|{\cal P}_{2f}\left|\Psi\right\rangle \approx 0$) 
except for the lowest $2^{+-}$ state where we obtained 
the opposite of the desired result for $(-1)^{\cal J}$. 

Although the overall fit with the conventional lattice
results is quite good, we see two deficiencies of
our spectrum that cannot be attributed to $K$ or $p$-truncation
errors.  First, we see that the lowest $0^{--}$ state is
too low in energy.  Second, we see that the lowest 
parity doublet $2^{\pm+}$ is not quite degenerate.
We believe that these discrepancies must be due to our truncation
of the effective potential.
Finally, we have made no prediction for the lowest $1^{++}$ state since
it lies too high in the $\left|{\cal J}\right|^{++}$ spectrum;
there may be a additional $0^{++}$ state of lower energy
which prevents us from labeling the $1^{++}$ unambiguously.

\begin{figure}
\centering
\BoxedEPSF{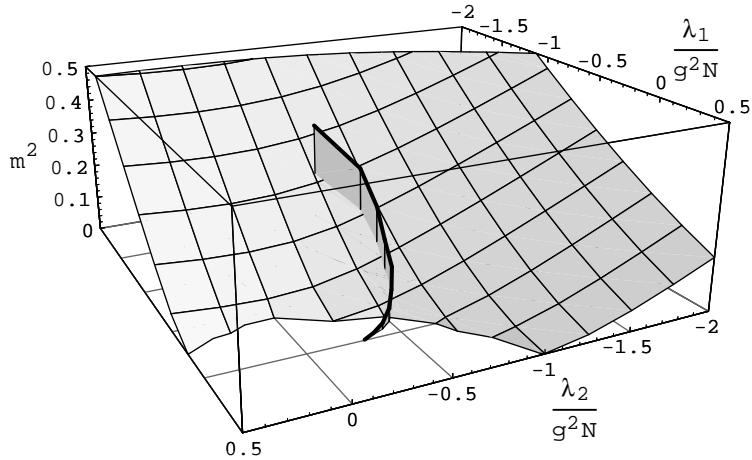 scaled 667}
\caption{ Mass $m^2= \mu^2 a/(g^2 N)$ such that the lowest $M^2$
eigenvalue is zero vs $\lambda_1/(g^2 N)$ and  $\lambda_2/(g^2 N)$
for $p \le 6$, $K=14$, $\lambda_3 =100 g^2 N$.  
Below this surface the spectrum is tachyonic and below $m^2=0$ our
quantisation breaks down.
The dark line is  the scaling trajectory. 
\label{fig3}}
\end{figure}

The numerical bound for absence of tachyons is shown in
Fig.~\ref{fig3} as a zero-mass surface.
The surface agrees rather well with the analytic bound provided by the
``transverse only'' model (\ref{bounds}) with $\gamma$ taken over the 
full range $0 < \gamma < 1$.
As the transverse lattice spacing vanishes the mass gap should vanish
in lattice units. 
The fixed point for this, which we believe lies somewhere at 
negative $m^2$, should lie on the zero-mass surface, but is inaccessible
to us in the dielectric regime $m^2 > 0$.
Nevertheless the scaling trajectory should gradually approach the
zero-mass surface if it is to eventually encounter the fixed point. 
We see this happening already in Fig.~\ref{fig3}.

%
%
%
%
\begin{table}
\begin{center}
\begin{tabular}{c|c|cccccc}
$\displaystyle {\left|{\cal J}\right|}^{{\cal P}_1 \cal C}$& 
$\displaystyle \frac{M^2}{\sigma}$ &
\bBoxedEPSF{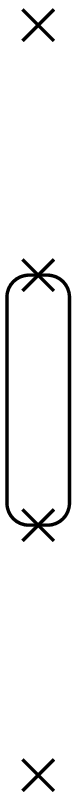 scaled 400}&
\bBoxedEPSF{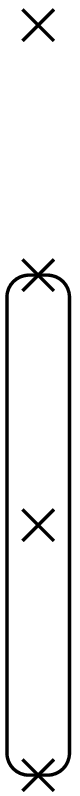 scaled 400}&
\bBoxedEPSF{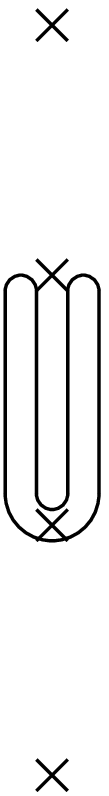 scaled 400}&
\bBoxedEPSF{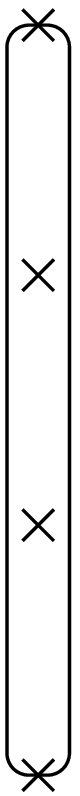 scaled 400}&
\bBoxedEPSF{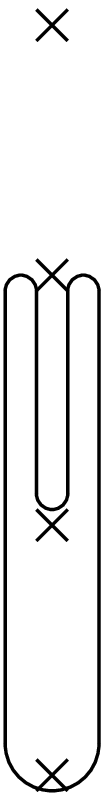 scaled 400}&
\bBoxedEPSF{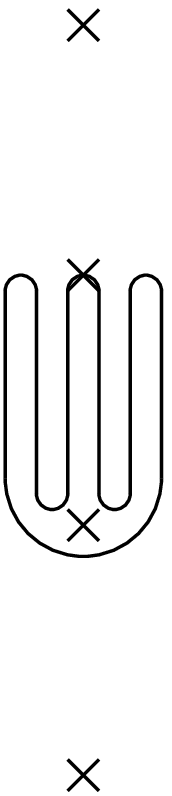 scaled 400}\\[10pt]
\hline
$0^{++}$ &  16.3 & 0.036 & 0.472 & 0.372 & 0.019 & 0.084 & 0.017 \\
$0^{++}$ &  36.2 & 0.046 & 0.359 & 0.528 & 0.019 & 0.024 & 0.023 \\
$0^{++}$ &  67.3 & 0.051 & 0.405 & 0.332 & 0.026 & 0.145 & 0.04  \\
$0^{--}$ &  27.3 & 0.909 & 0.075 & 0.011 & 0.002 & 0.002 & 0     \\
$0^{--}$ &  50.5 & 0.06  & 0.681 & 0.153 & 0.037 & 0.065 & 0.003 \\
$0^{--}$ &  73.5 & 0.02  & 0.16  & 0.741 & 0.014 & 0.048 & 0.018 \\
$0^{+-}$ &  109. & 0     & 0.142 & 0.026 & 0.026 & 0.801 & 0.005 \\
$0^{-+}$ &  86.3 & 0     & 0.244 & 0.671 & 0.01  & 0.059 & 0.016 \\
$2^{++}$ &  38.8 & 0.86  & 0.09  & 0.033 & 0.006 & 0.011 & 0     \\
$2^{++}$ &  82.4 & 0.022 & 0.113 & 0.096 & 0.078 & 0.512 & 0.179 \\
$2^{-+}$ &  51.8 & 0     & 0.885 & 0.014 & 0.027 & 0.074 & 0     \\
$2^{-+}$ &  94.9 & 0     & 0.16  & 0.036 & 0.011 & 0.793 & 0.001 \\
$2^{--}$ &  67.1 & 0.859 & 0.082 & 0.05  & 0.003 & 0.005 & 0.001 \\
$2^{--}$ &  102. & 0.054 & 0.609 & 0.132 & 0.061 & 0.136 & 0.007 \\
$2^{+-}$ &  67.  & 0     & 0.909 & 0.003 & 0.037 & 0.051 & 0     \\
$2^{+-}$ &  124. & 0     & 0.701 & 0.126 & 0.027 & 0.142 & 0.004 \\
$1^{-+}$ &  92.2 & 0     & 0.549 & 0.156 & 0.022 & 0.269 & 0.003 \\
$1^{--}$ &  79.4 & 0.047 & 0.398 & 0.447 & 0.013 & 0.081 & 0.014 \\
$1^{+-}$ &  84.4 & 0     & 0.895 & 0.007 & 0.034 & 0.063 & 0   
\end{tabular}
\end{center}
\caption{The transverse structure of the spectrum for
$K=14$, $p \le 6$.  The first column is our assignment 
of ${\left|{\cal J}\right|}^{{\cal P}_1 \cal C}$,
the second column is the extrapolated $(\mbox{mass})^2$ from 
Fig.~\protect\ref{fig4}, and the remaining
columns show the probability for a state to have a given 
transverse structure.  
The coupling constants are from the $m^2=0.065$ row of 
Table~\protect\ref{scaling}.
\label{looper}}
\end{table}

Looking at the transverse structure  in Table~\ref{looper}, one
can see several important properties of the spectrum.
First we notice that the lowest $0^{++}$ state has very 
little two particle content.
This is the main effect of the 
$\left(\Tr\left\{M^\dag M\right\}\right)^2$ interaction
when the associated coupling constant $\lambda_3$ is large.
It essentially removes the two particle sector from the
wavefunction of this state.
Second, we notice that there is a relatively large
mixing between the various sectors, including mixing of sectors of 
different particle number at the 10-20\% level.  This is large enough
to restore some rotational invariance but not so large that it
contradicts our truncation of the effective potential.
If we look at parity doublets, we see that each   
wavefunction is quite dissimilar from its partner, including very different
particle content.  This is what one would expect since 
full parity is dynamically complicated in light-cone coordinates.
Finally, one would expect from Regge trajectories
that the lowest $2^{++}$ state would have
a large two particle content while the first excited 
$0^{++}$ state should not.  This helps us label the correct
$\left|{\cal J}\right|$ for these two nearly degenerate 
states.  In fact, large mixing between these states may explain
why we do so poorly with the lowest $2^{{\cal P}_1 +}$ parity doublet.

\subsection{Is our spectrum accidental?}
\label{question}

Since the scaling trajectory in Table~\ref{scaling} is
a {\em local}\/ minimum of $\chi^2$ in coupling constant space,
one might worry that our agreement with the ELMC data is
accidental.    Since we are not always able to assign $|{\cal J}|$
to our states, it is possible that another local minimum
in parameter space represents the `correct' scaling 
trajectory.  With this in mind, we have conducted a thorough search
of parameter space and have found two other trajectories.

The first occurs for $\lambda_2 \approx 0.5 g^2 N$
and $\lambda_1 <0$.  However, the $\chi^2$ of the fits are 
significantly worse.  Moreover, it goes against our theoretical
prejudice that the scaling trajectory should be in the vicinity of,
and track towards, the zero-mass surface (fig.(\ref{fig3})).

The second occurs quite close to our scaling trajectory.
We have included $\lambda_1$ and $\lambda_2$ for this 
trajectory as dashed lines in Fig.~\ref{fig12}.  
Indeed, the $\chi^2$ of the
fits are as good or better than that of our chosen trajectory.
We find that the principle difference between the two trajectories
is that the $2^{++}$ and $0^{++}$ levels are switched --- recall that
we are unable to distinguish these levels by exact symmetries alone.
Consistent with the Regge trajectory picture, we believe
that two particle states should form the lowest Regge trajectory
and hence the lowest $2^{++}$ state should contain mainly
two particles.  Thus, we have chosen the scaling trajectory where
the lowest $2^{++}$ state has mainly two particle content.
Additionally, we find that the ratio 
$(g^2 \sigma)_{\rm \LCTL{}}/(g^2 \sigma)_{\rm ELMC}$
in Fig.~\ref{fig5} 
is not as good for this other trajectory (dashed line), nor does it appear
to track towards the zero-mass surface.
%

\section{The String Tension.}
\label{sec5}

To measure the string tension in the $x^1$ direction 
(before Eguchi-Kawai reduction)
consider a lattice with $n$ transverse links and periodic
boundary conditions. Constructing a basis of Polyakov loops,
``winding modes,'' that wind once around this lattice, one
may extract from the lowest eigenvalue $M^2$ vs $n$ 
the lattice string tension $a\sigma = \Delta M_n/\Delta n$.
Because of Eguchi-Kawai reduction, this is equivalent to using
Polyakov loops of winding number $n$ on the single-link periodic lattice
(Eqn.~(\ref{wf}) with $n \neq 0$).

\subsection{Analytic results}

The behavior of Eqn.~(\ref{vanish}) is equally true of winding modes
$n \neq 0$.  Thus, it is useful to extend the ``transverse only'' model
of Section~\ref{transverseonly} to the case of nonzero
winding number $n$.
In particular, consider the $p=n$ particle state 
$\left|+1, +1, \cdots +1\right\rangle$
and the $p=n+2$ particle state $\left|-1, +1, +1,  \cdots +1\right\rangle$.
The correct normalisation of these states is 
\begin{eqnarray}
    1 &=&\left\langle +1, +1, \cdots +1|+1, +1, \cdots +1\right\rangle 
      = p \, C_{+1,+1,\ldots,+1}^2\\
     1 &=&\left\langle -1,+1, +1, \cdots +1|-1,+1, +1, \cdots +1\right\rangle 
      = C_{-1,+1,+1,\ldots,+1}^2 \; .
\end{eqnarray}
Using these two states as a basis, we obtain a $2 \times 2$ mass 
squared matrix $(n>1)$
\begin{equation}
  \frac{1}{4 a} \left( \begin{array}{cc}
    \pi n(n-1)\left(g^2 N + \lambda_2 \right) &
    4 n\sqrt{n+1} (\lambda_1 + \lambda_2) \\
    4 n\sqrt{n+1} (\lambda_1 + \lambda_2) & 
    \pi(n+1)(n+2) \left(g^2 N + \lambda_2 \right) 
    +4 \pi (n+1) \lambda_1 
  \end{array} \right)+ O(\beta)  \; . \label{mass}
\end{equation}
The diagonal elements come from Eqn.~(\ref{tro}), setting 
$\gamma=1$ and $\gamma=(n+1)/(n+2)$, respectively.
The off-diagonal matrix element is, more precisely,
\begin{eqnarray}
   \lefteqn{\left\langle+1, +1, \cdots +1|P_\mu P^\mu|-1,+1, +1, \cdots +1\right\rangle}
    \nonumber \\ &=&
    \frac{\left(\lambda_1+\lambda_2\right)
    \Gamma(4 \beta+1)\, {\Gamma\!\left(\beta+\frac{1}{2}\right)}^3
         \sqrt {n \, \Gamma\left((2 \beta+1)n\right)
            \, \Gamma\left((2 \beta+1)(n+2)\right)}}
         {2 \pi a \, \Gamma\!\left(3 \beta+\frac{3}{2}\right) 
                  {\Gamma\left(2 \beta+1\right)}^2 \,
               \Gamma\left(2 \beta (n+1)+n\right)} \; .
\end{eqnarray}
An example of the lowest eigenvalue of Eqn.~(\ref{mass}) vs $n$
is plotted in Fig.~\ref{fig1}.
For large $n$ one obtains a linear spectrum 
$M\approx n\sqrt{\left(\lambda_2 + g^2 N\right)/a}$
and the vanishing lattice string tension 
condition $\lambda_2 = - g^2 N+O(\beta)$. 
%
%
%
%
%

\subsection{Numerical results}

%
\begin{figure}
\centering
\BoxedEPSF{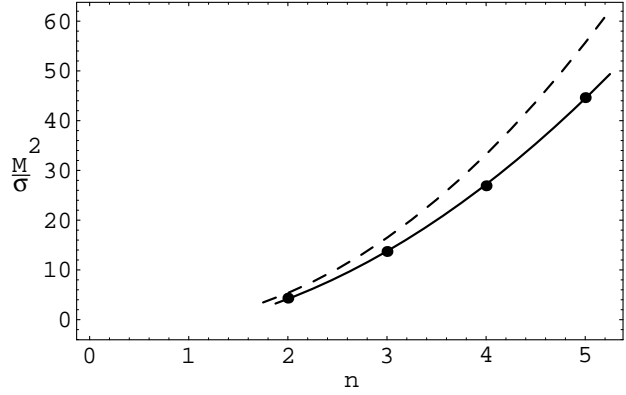 scaled 667}
\caption{Lowest $M^2$ eigenvalue vs 
$n$ for winding modes.
Here, $K=10.5$ or 11, $p \le n+4$, and the couplings 
are taken from the $m^2 =0.065$ row of Table~\protect\ref{scaling}.
Also shown is a numerical fit to $1.92 n^2-3.5$.
The dashed line is from the variational estimate of
Eqn.~(\protect\ref{mass}). There is a subtlety associated with the $n=1$
point, which is described in the text.
 \label{fig1}}
\end{figure}

%
%
\begin{table}
\begin{center}
\begin{tabular}{c|cc}
$\displaystyle m^2=\frac{\mu^2 a}{g^2 N}$ & 
$\displaystyle a^2 \sigma$ & 
$\displaystyle \frac{b}{\sigma}$ 
\\[10pt]
\hline
 0.      & 1.48 & -2.77  \\
 0.00394 & 1.56 & -2.95 \\
 0.0158  & 1.65 & -3.08   \\
 0.036   & 1.77 & -3.29  \\
 0.065   & 1.92 & -3.5   \\
 0.104   & 2.09 & -3.68  \\
 0.153   & 2.3  & -3.82  \\
 0.214   & 2.55 & -4.06  \\
 0.291   & 2.86 & -4.28  \\
 0.384   & 3.29 & -4.21 
\end{tabular}
\end{center}
\caption{String tension measurements along the 
scaling trajectory.  The coupling constants
are taken from Table~\protect\ref{scaling}.  Here,
the lowest winding mode eigenvalue $M^2_n/\sigma$ has been fit to
$a^2 \sigma n^2+b/\sigma$ for $n=2,3,4,5$.  
\label{tentable}}
\end{table}

Numerical calculation of the winding modes proceeds in the same
manner as our calculation of the spectrum except that we use a basis
of $n\neq 0$ states.
String theory arguments indicate that oscillations of the winding 
mode transverse to itself yield  a form 
$M^2 = \sigma^2 (na)^2 - \sigma \pi /3$, the constant correction 
being due to Casimir energy~\cite{luscher}. 
Fig.~\ref{fig1} shows a typical $M^2$ vs $n$ plot for winding modes.
Consistent with the expectations from string theory,
we have fit the winding mode spectrum to a curve of the form
\be
     \frac{M^2_n}{\sigma} = a^2 \sigma n^2 + \frac{b}{\sigma}\; , \label{nnn}
\eq
which we expect to be appropriate for QCD at sufficiently large $n$.
As illustrated by Fig.~\ref{fig1}, there is a good fit to the
quadratic. 
%
%
In the same manner, we determine (\ref{nnn}) all along
the scaling trajectory with the result is shown in Table~\ref{tentable}.

The constant term $b/\sigma$ represents the leading finite
transverse size correction to the string tension, and is supposed to
be universal. The fact that it
varies strongly along our `scaling trajectory', shows that for
this quantity there are still large finite $a$ errors. Also we have not
extrapolated in $K$ or $p$ truncation. 
There is a subtlety associated with the $n=1$,
one particle matrix element
\be
   \frac{1}{N} \left\langle 0 \right| \Tr\left\{a_{1}(K)\right\} P_\mu P^\mu
        \Tr\left\{a^\da_{1}(K)\right\}\left|0\right\rangle \; . \label{onep}
\eq
The most consistent definition is to assume that this matrix element is
infinite due to the diagonal part of the $J \left(\partial_-\right)^{-2} J$
interaction.  
In this case,
 $M^2_{n=1}$ lies somewhat above
the extrapolation of the solid line in Fig.~\ref{fig1} down to $n=1$,
since the eigenfunction contains only $3,5,7,\ldots$--particle 
contributions.
(If one instead defines (\ref{onep}) to be equal to $\mu^2$,
$M^2_{n=1}$ falls right along the extrapolation of the quadratic fit.)
We have not included the $n=1$ datum in our determination of the
string tension.

\begin{figure}
\centering
\BoxedEPSF{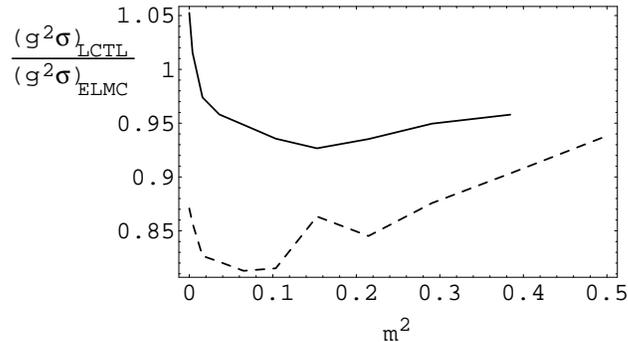 scaled 667}
\caption{The ratio $(g^2 \sigma)_{\rm \LCTL{}}/(g^2 \sigma)_{\rm ELMC}$.
The solid line is for our scaling trajectory
while the dashed line is from the second trajectory discussed in
Section \ref{question}.
 \label{fig5}}
\end{figure}

If we are careful to distinguish between our measurement of $g^2 N$ 
and $\sigma$ versus $g^2 N$ and $\sigma$ used in the ELMC
calculations, we can form the ratio
\be
   \frac{ \left(g^2 \sigma \right)_{\rm \LCTL{}}}
     { \left(g^2 \sigma \right)_{\rm ELMC}}
    = \left(\frac{ \sqrt{\sigma}}{g^2 N}\right)_{\rm ELMC}
    \cdot \frac{\left(g^2 N\right)_{\rm \LCTL{}}}
         {a \, \sigma_{\rm ELMC}}
    \cdot \sqrt{\frac{a^2 \sigma^2_{\rm \LCTL{}}}
          {\sigma_{\rm ELMC}}}
\eq
%
%
The first term is $0.1974(12)$ according to Teper's 
large $N$ extrapolated measurement
of $\sqrt{\sigma}/(g^2 N)$ and the remaining two terms are 
{}from Tables \ref{scaling} and \ref{tentable}, respectively.
This ratio measures our error in measuring the string tension
plus any discrepancy between $g^2 N$ as used by us and that
used in the ELMC calculation.  From Fig.~\ref{fig5}, we see 
that the two couplings differ by 5\%.  This implies
a rather large discrepency since we
have used a non-standard definition of $g^2$.

In general, we find that the surface in our parameter space on which 
$M^2$ vanishes (in units of $g^2 N /a$) , Fig.~\ref{fig3}, 
always lies above the surface of vanishing lattice string
tension $a^2 \sigma$; that is, the spectrum always becomes tachyonic before
the lattice string tension vanishes.  
Any ``critical point'' must occur in the intersection
of the surface of vanishing lattice string tension and vanishing mass gap.  
In fact, we find no such point
anywhere in the $m^2>0$ coupling constant space, in agreement with the
colour dielectric arguments of Section~\ref{motivate}. Instead we have a scaling
trajectory at finite lattice spacing, as we now show.

%

\begin{figure}
\centering
\BoxedEPSF{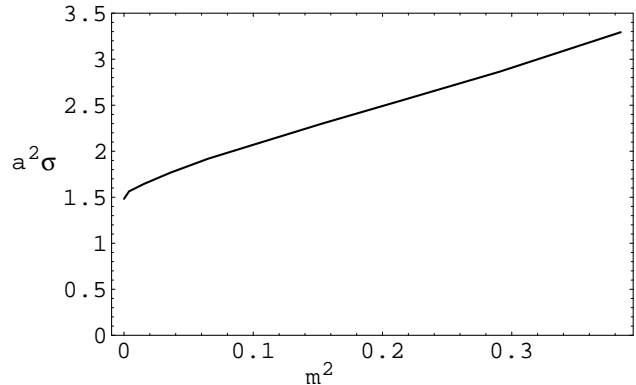 scaled 667}
\caption{The lattice spacing in units of the string
tension along the scaling trajectory, Table~\protect\ref{tentable}.  
Here we plot $a^2 \sigma$ vs $m^2$.
 \label{fig9}}
\end{figure}

If we assume that our $\sigma$ is equal to the ELMC value of
$\sigma$, we can determine the lattice spacing $a$ in units
of the string tension along the scaling trajectory.  
As mentioned previously, this
demonstrates that the mass $m^2=\mu^2 a/(g^2 N)$ determines 
the lattice spacing and that the continuum limit
occurs at $\mu^2 <0$; see Fig.~\ref{fig9}.

%
%
%

\section{Density of States and $T_c$.}
\label{sec6}

Another quantity of interest is the deconfinement temperature $T_c$,
which on general grounds one expects to exist in a pure gauge theory
in greater than $1+1$ dimensions \cite{svet}. In the Euclidean finite
temperature formalism, this is signalled by a
winding mode in the Euclidean time direction which becomes
tachyonic. An estimate of the inverse circumference $T=T_c$ at which 
this occurs comes from the  formula 
\cite{stringy}
for the lowest mass eigenstate of
a long closed harmonic string of winding number one:
\be
M^2 = {\sigma^2 \over T^2} - {\sigma \pi (D-2) \over 3}
\eq
where $D$ is the number of spacetime dimensions. It becomes tachyon at
\be
T_c = \sqrt{3 \sigma \over \pi (D-2)} \label{tcrit}
\eq
The result $T_c \approx 0.98 \sqrt{\sigma}$ for $D=3$ 
is  close to, but different from, 
that predicted by the large-$N$ extrapolation of the
ELMC results \cite{teper}:\footnote{The numerical ELMC result is
in turn in agreement with the analytic large-$N$ result of the
Torino group for $T_c$ based also on Euclidean lattice methods \cite{turin}.}
   $T_c = 0.89(3) \sqrt{\sigma}$. 

$T_c$ may also be obtained from the Hagedorn behaviour \cite{hage}
 of the asymptotic density of mass eigenstates
\be
\rho(M) \sim  M^{-\alpha} \exp (M/T_c) \label{density}
\eq
The canonical partition function diverges for $T>T_c$. If $\alpha >
(D+1)/2$ it is a phase transition, beyond which the canonical and 
microcanonical ensembles are inequivalent. If $\alpha < (D+1)/2$ the
ensembles are equivalent and $T_c$ represents a limiting temperature
--- the free energy diverges at $T_c$.
The light-cone quantisation of a free relativistic string with
harmonic oscillations in the transverse direction (Nambu-Goto Bosonic
string)
was first performed by Goldstone {\em et alii} \cite{string}, who found a
linearly rising discrete spectrum of squared masses
\be
M^2 \sim 2 \pi \sigma n \label{regge} \; , \; \; \; \; n \in Z \label{gold}
\eq
very similar to
the hadronic Regge trajectories observed in nature. Each mass level
has a degeneracy increasing with $n$, such that the density of
states $\rho$ per unit mass yields Eqn.~(\ref{tcrit}) once more.
Thus large-$N$ QCD seems to have slightly larger  level density than this
Bosonic string. 

In the next section, we will use the density $\rho(M)$ in the
\LCTL{} formulation to understand analytically this difference
between $T_c$ in the large $N$ gauge theory and the Bosonic string.
We will discuss numerical estimates of $T_c$ from the 
Hagedorn behavior in Section~\ref{numest}. 

\subsection{Analytic Regge Trajectories.}

High mass glueballs are not quite described by the Bosonic string model.
One piece of evidence for this is the discrepancy in $T_c$ noted above. By
construction, the only degrees of freedom, in addition to the centre
of mass, allowed in the Bosonic string model leading to 
(\ref{tcrit}) are transverse
oscillations. A `QCD string' might also be expected to have longitudinal
degrees of freedom. In fact it is well-known that allowing a
relativistic string to oscillate only in transverse directions leads
to consistency problems, for example with Lorentz invariance, unless $D=26$. A
correct treatment when $D \neq 26$ must allow longitudinal
oscillations in an essential way \cite{poly}. Recent attempts, by one
of the authors, to study directly the light-cone quantisation of such
`non-critical' string models have led to largely
inconclusive results however \cite{nonc}.
The Wilson loops of our light-cone lattice dielectric theory clearly
exhibit oscillations in the both the longitudinal $x^-$ and transverse
$x^1$ directions. We propose that the former accounts for the
slight shift in $T_c$. 

It is instructive to derive the Hagedorn behaviour in the \LCTL{}
approach and show how the 
extra degrees of freedom necessary to lower
$T_c$ from its Bosonic string value
might arise.
The exponential density of asymptotically high mass eigenfunctions of 
$P^-$ (\ref{energy}) derives from wavefunctions containing typically
many links $p$, i.e. long strings. 
It is useful to consider two extremes of this large
$p$ problem, according to the complexity
in the longitudinal dependence of the wavefunctions $f$ on Bjorken
$x=k/P^+$, conveniently measured by the typical number of nodes say.
We split $P^- = P^{-}_{t} + P^{-}_{l}$ into transverse and
longitudinal pieces,
where from Eqn.~(\ref{energy})
\begin{eqnarray}
P^{-}_{t} &  = &   \int dx^-\, V[M]  \\
P^{-}_{l} &  = &   -\int dx^- \, {g^2 \over 2a} \Tr \left\{ 
      J^{+} \frac{1}{\partial_{-}^{2}} J^{+} \right\}
\end{eqnarray}
There are two problems which face us now. Firstly, we expect that more
terms must be added to $V[M]$ as we examine levels of higher mass in
order to restore Lorentz invariance for them. Secondly,
the wavefunction will not in general split into a
product of longitudinal and transverse pieces. There are two extremes
when one might expect a degree of universality to help.

Consider first the extreme when $p$ takes its maximum value, for
example when each link variable assumes the minimum $x=1/K$ allowed
by the DLCQ cut-off. 
The solutions of $P^{-}$ in this limit
have already been studied by Klebanov and Susskind \cite{ks} (see also
Ref.~\cite{dalley1}) in the approximation that processes which lower
the number of particles are neglected.\footnote{We will not
regurgitate the somewhat technical analysis,
but refer the interested reader to the original literature.} 
They found that for a wide range of parameters in the
potential $V$ the spectrum was precisely (\ref{regge}) with the
correct
degeneracy to produce (\ref{tcrit}). $P^{-}_{l}$
vanishes in the approximation, so one essentially diagonalises
$P^{-}_{t}$. They also argued that allowing some particle annihilation
and creation would merely renormalise $\sigma$ to a first approximation.
This would mean that, 
as far as the asymptotically high spectrum is concerned, the
purely transverse degrees of freedom of glueballs in the \LCTL{} 
formulation are just those of the Bosonic string.

\begin{figure}
\centering
\BoxedEPSF{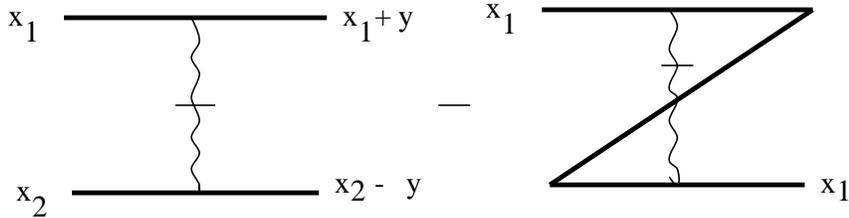 scaled 667}
\caption{Coulomb exchange diagram and its self-energy
contraction. Thick lines are propagating particles, wavey line the
non-propagating $A_+$ mode. \label{coulomb}}
\end{figure}

In the previous extreme the longitudinal dynamics were frozen by
construction. Let us now unfreeze them by allowing oscillations
in the longitudinal direction, with non-trivial dependence of the
wavefunctions on Bjorken $x$, and go to the other extreme.
The high spectrum for $P^{-}_{l}$ typically contains wavefunctions
$f_{\lambda\rho\ldots\sigma} (x_1, \ldots, x_p)$ 
which oscillate rapidly as a function of each $x_i$. In this case
we can make the asymptotic approximation suggested by 't Hooft \cite{thooft}
for fundamental representation fermions in two-dimensional light-cone
large-$N$ QCD; in fact our initial 
analysis is only a slight modification of the
corresponding one for adjoint representation fermions \cite{dalley2} 
due to Kutasov \cite{kut}. For such 
rapidly oscillating wavefunctions all of the
terms in $P^{-}_{t}$ tend to average to 
zero. The same is true of the terms in $P^{-}_{l}$, except for the 
Coulomb exchange term which is
singular at zero momentum transfer ($k_2 = k_3$ in Eqn.~(\ref{coul})),
and  the
creation of small-$x$ particles from the same
instantaneous
$A_+$ mode (Eqn.~(\ref{create}) at $k_1 + k_2 =0$ or $k_2 + k_3 =0$).
Keeping only the Coulomb exchange term for the moment, 
the $P^{-}_{l}$ eigenvalue problem is dominated by the following
part of the light-cone Schr{\"o}dinger equation
for the wavefunction components of a $p$-link state
\begin{eqnarray}
M^{2} f_{\lambda\rho\ldots\sigma} (x_1, \ldots, x_p)
&  = & {g^2 N \over \pi a}  \int_{-\infty}^{\infty}
{dy \over y^2} \{ f_{\lambda\rho\ldots\sigma} (x_1, \ldots,
 x_p) \nonumber \\ && - f_{\lambda\rho\ldots\sigma} (x_1+y, x_2
 - y, \ldots, x_p) \}   + \; \mbox{cyclic permutations}
\label{eig}
\end{eqnarray}
This is obtained by projecting the Coulomb part of
$P^{-}_{l} \left|\Psi \right\rangle $ onto individual
Fock states. The Coulomb kernel has been simplified and the limits of the
integral extended to the whole real line under the assumption that
the dominant contribution comes from $y \approx 0$. The right hand
side
of (\ref{eig}) is 
represented by the diagrams of light-cone perturbation theory in 
Fig.~\ref{coulomb}.
We must decide what to do about the transverse shape of the Wilson
loop, represented by the 
indices
$(\lambda,\rho,\ldots,\sigma)$. 
In principle there will be a fine
structure spectrum of transverse shapes on top of the large mass eigenvalue
provided by Eqn.~(\ref{eig}). We will make the approximation that this
can be represented by an overall transverse degeneracy factor,
which for simplicity we will take as the number of transverse
configurations of a $p$-link loop $\sim 2^p$. This is an overcounting,
obviously including lattice artifacts, and a more sophisticated
treatment is no doubt possible. With the random distribution of
transverse
configurations, we can suppress the indices $(\lambda,\rho,\ldots,\sigma)$
for the following; however, the number of links $p$ is remains significant. 
Solutions to (\ref{eig}) must satisfy the additional boundary conditions
\be
f(x_1, x_2, \ldots, x_p) = 0  \; \mbox{if any} \;  x_m =
0 \; , \; \;  m \in \{1, \ldots ,p\} \; . \label{drop}
\eq
If we look for solutions symmetric under a cyclic
permutation of the $x_m$'s, a sufficient condition for (\ref{drop})
is that $f(0, x_2, \ldots, x_p) = 0$. 
The solutions are essentially those given by Kutasov \cite{kut} with 
minor modifications for Boson statistics. We just display the
first couple:
\be
f(x_1,x_2) = \sin{\pi n_1 x_1} \; ; \; \; \; \; M^{2} = 2g^2 N \pi n_1/a  
\; ; \;\;\;\; \mbox{$n_1$ odd integer} 
\eq
\begin{eqnarray}
f(x_1,x_2,x_3,x_4) & = &  \sin{\left(\pi n_1 (x_1 + x_2)\right)} 
  \sin{\left(\pi n_2 \left(x_2 +
x_3\right)\right)} - \sin{\left(\pi n_1 x_1\right)} 
\sin{\left(\pi n_2 x_3\right)} \nonumber \\
 && + \sin{\left(\pi n_1 x_2\right)}\sin{\left(\pi n_2 x_4\right)} + 
(n_1 \leftrightarrow n_2)\; ; \\
M^{2} & = & 
2 g^2 N \pi (n_1 + n_2) /a \;  ;  \; \; n_1 > n_2 \; ; \;\;\;\;
  \mbox{$n_i$ odd integers.} \nonumber
\end{eqnarray}
The general spectrum is 
\be
M^{2} = {2g^2 N \pi \over a} (n_1 + n_2 + \cdots + n_{p/2}) \; ;
\;\;\;\; \mbox{$n_i$ odd.} \label{general}
\eq

\begin{figure}
\centering
\BoxedEPSF{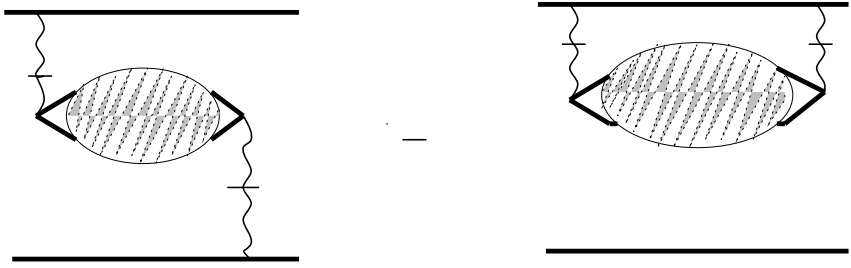 scaled 667}
\caption{Self-energy corrections to the non-propagating $A_+$
mode. These diagrams represent  dominant particle non-conserving
processes when small momentum fraction
$y$ is flowing through the initial $A_+$ line.
\label{self}}
\end{figure}

In general the high mass states are a mixture of different numbers of particles
$p$, so Eqn.~(\ref{eig}) cannot be the whole story. We must also take
into account pair creation effects. The most important effect is likely to
be the $A_+$ self-energy processes Fig.~(\ref{self}), which are singular
when small momentum fraction $y$ is flowing through the instantaneous $A_+$ 
line.
The net effect of such ``vacuum polarisation'' is to give a 
renormalisation of the Coulomb coupling constant $g^2 N/\pi a$ and hence the 
mass scale setting for Eqn.~(\ref{general}). In fact such a
renormalisation must occur since the bare expression (\ref{general})
is obviously  $a$ dependent; the Coulomb exchange process cannot
produce any motion on the transverse lattice. 
Through pair creation and annihilation of links, transverse
motion becomes possible. We will not need  to determine the
renormalisation for the present application, since the
spectrum is still of the string theory form (\ref{gold}) and our ignorance
of the functional dependence of the mass scale on the couplings 
is parameterised by $\sigma$.

{}From the form (\ref{general}) 
we may calculate the density of states using standard
methods of statistical mechanics. The generating function
\be
G(w) = \sum_{n} d_n w^n
\eq
defines the number of states $d_n$ of $M^{2} = 2 \pi \sigma n$. 
For large $n$ we need to find the behaviour as $w \to 1$. Including 
the transverse degeneracy factor $2^p$ 
we have explicitly
\begin{eqnarray}
\log{G(w)} & = & \log{\prod_{m=1}^{\infty} (1 + 4w^m)} \nonumber \\
     & = &   -\sum_{m,q=1}^{\infty} {(-4w^m)^q \over q} \nonumber \\
     & = &  -\sum_{q=1}^{\infty} {(-4w)^q \over q (1-w^q)} \nonumber \\
     & \sim & - {1 \over 1-w} \sum_{q=1}^{\infty} {(-4)^q \over q^2} 
           \;\;\;\; (w \sim 1) \nonumber \\
     & = & -{ {\rm Li}_{2}(-4) \over  (1-w)}
\end{eqnarray}
where ${\rm Li}_{2}$ is the dilogarithm function. Then
\be
d_n = {1 \over 2\pi i}\int_{\cal C} dw \, {G(w) \over w^{n+1}} \label{cont}
\eq
where ${\cal C}$ encircles the origin. The integral (\ref{cont}) has a
saddle at $\log{w} = \sqrt{-{\rm Li}_{2}(-4)/n}$ giving
\be
d_n = {\rm exp} \left[ 2 \sqrt{-n {\rm Li}_{2}(-4)} \right] \approx 
{\rm exp} \left[ M \sqrt{4.75 \over \pi \sigma} \right] \label{den}
\eq
The Hagedorn temperature $T_c = 0.81 \sqrt{\sigma}$ should  be
regarded as a lower bound given the probable overestimate of the
degeneracy of transverse configurations. The important point to note
is that the density of states would rise exponentially as a result
of longitudinal excitations alone,  even if we did not include any
degeneracy factor for  transverse structure. This
is strong evidence that $T_c$ can be lower in large-$N$ gauge theory than
in free Bosonic string 
theory as a result of non-trivial longitudinal degrees of
freedom. Our estimate, therefore, is  that it lies in the
range $0.81 \sqrt{\sigma} < T_c <  0.98 \sqrt{\sigma}$,
consistent with the ELMC result.

\subsection{Numerical estimates}
\label{numest}

%
%

%
\begin{figure}
\centering
\BoxedEPSF{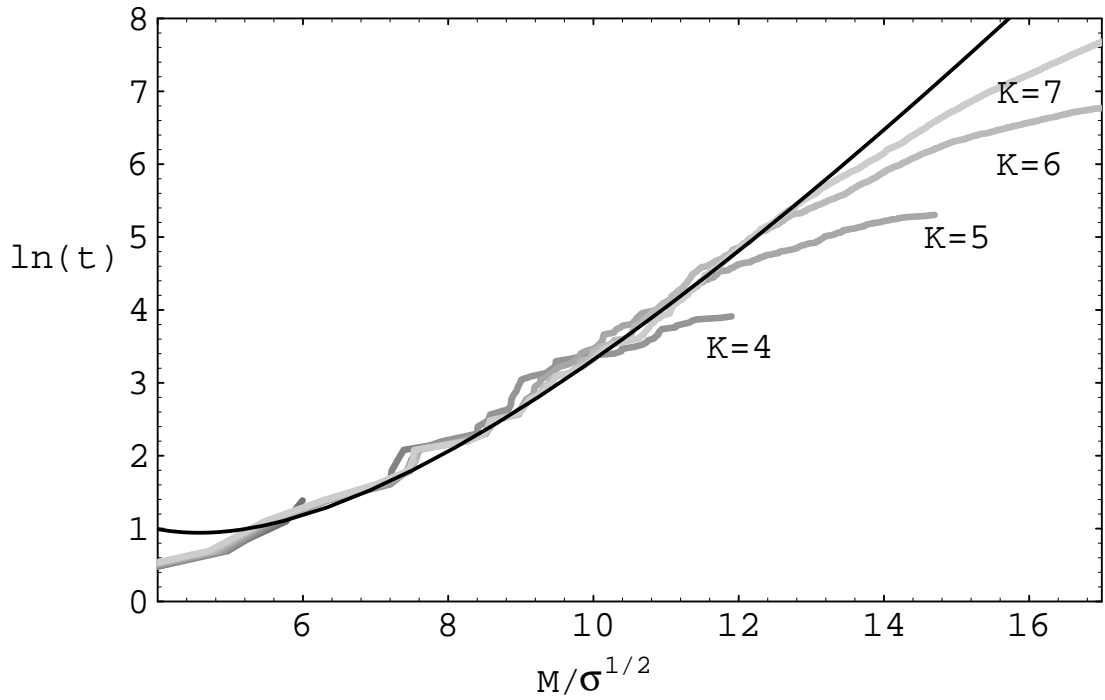 scaled 1000}
\caption{
$\ln t$ vs the mass of the $t$-th eigenvalue $M_t$ in
units of the string tension.  Here we have applied 
a cutoff in $K$ only.  Also shown is a least squares fit to 
the $K=7$ data in the range $0.5<\log(t)<5.5$.
We have chosen the couplings from the $m^2=0.065$ 
line of Table~\protect\ref{scaling}.
\label{fig21}}
\end{figure}

Now we turn our attention to a numerical estimation of the 
density of states.  For this calculation, we have only specified
the box size $K$ to truncate the basis.  
It is essential to demonstrate that the spectrum is sufficiently
converged in $K$, thus we only fit to the states between 
$0.5<\log(t)<5.5$ for the $K=7$ data; see Fig.~\ref{fig21}.  
A numerical fit gives,
%
\be
      \log(t) =3.99 + \frac{M}{0.78\sqrt\sigma} - 
             5.86 \log\left(\frac{M}{\sqrt\sigma}\right) \; .
\eq
Since the density of states is $\rho(M)=dt/dM$,
we find  $T_c=0.78 \sqrt\sigma$ with an estimated 
error\footnote{This result could be 
easily improved upon since the $K=7$ truncation contains only 4117 states.}
of  at least 10\%. Due to the large error, this result is both compatible
with the Euclidean lattice result \cite{teper,turin} 
and the analytic bound of the previous subsection which assumed
random transverse configurations.
The fact that the power correction 
$\alpha \approx 4.86$ 
%
%
in the
density (\ref{density}) is much larger than $(D+1)/2 = 2$
means that the thermodynamic free energy
remains finite at $T_c$, marking a true phase transition (presumably
deconfinement) rather than a  limiting temperature.

\section{Conclusions}
\label{conclude}

We performed light-cone quantisation of  the colour 
dielectric lattice gauge theory on a transverse lattice, for $2+1$
dimensions and large $N$. The glueball solutions of the theory were
investigated as a function of couplings in an effective potential 
$V[M]$ on the transverse lattice, which contained all possible
gauge-invariant terms up to 4th order in $M$; 4 couplings in total
(plus an overall mass scale). We found an  approximate  scaling 
trajectory --- a one-parameter set of couplings for which physical
results are almost unchanged --- along which our glueball
spectrum agreed quite well with results from Euclidean lattice Monte Carlo
simulations; some 18 or 19 energy levels. 
The surface of vanishing mass in lattice units was also found in the region of
parameter space where our quantization is valid. 
Our scaling trajectory correctly
moves towards this surface as we decrease the lattice spacing. 
It presumably
differs from the exact scaling trajectory by higher order terms in
$V[M]$ that we neglected and which represent at this stage a
systematic error.
However, these corrections are seen to become less important as one
moves along the trajectory toward smaller lattice spacing.
We also bounded and estimated the deconfinement temperature
{}from the density of asymptotically high masses,
suggesting that  
longitudinal string-like oscillations are responsible for  deviations
{}from the Nambu-Goto string model.
Our methods in addition yield the light-cone wavefunctions for every
glueball in terms of the dielectric link variables, a more detailed
analysis of which will be presented in future work.
We applied  new
numerical methods for the solution of light-cone Hamiltonians which 
should be useful for a range of other problems.
Below we mention a couple of important questions arising from  our 
analysis and
discuss improvements and extensions to this work that we are carrying out.

\subsection{Large-$N$ phase transitions?}
\label{phase}
In lattice gauge theories with a simple action, such as the Wilson
action, there is a strong crossover behaviour between
strong and weak coupling which  probably turns into a
phase transition in the large $N$ limit. 
This large $N$ phase transition is well established in exactly solvable
models \cite{gross} and is a lattice artifact, 
{\em id est} it occurs even on a single link. (Remember there are an
infinite number of degrees of freedom per link at large $N$.)
We have been tacitly assuming throughout this work  that the infrared
stable RG trajectory does not encounter such a large-$N$ obstacle
between the large lattice spacing regime (where the
dielectric approach is valid) and the infra-red fixed point. 
Any truncation of the effective
potential $V[M]$ almost certainly will encounter a large-$N$
transition. This is simply illustrated by the following exactly
solvable one-link theory \cite{sdetal}
\be
Z = \int  {\cal D}M\, \exp \left( - \mu^2 \Tr\left\{M^{\dagger}M\right\} -
\Tr\left\{W(M^{\dagger} M)\right\} \right)
 \label{simple}
\eq
where $W(x) = g_2 x^2 + g_3 x^3 + \cdots $ is a {\em generic}
polynomial potential,
with all $g_i > 0$ (this condition is convenient but not essential). 
Decomposing $M = U^{\dagger} D V$ into a real diagonal matrix $D$ and 
unitary matrices $U$ and $V$, the solution can be expressed entirely in
terms
of the distribution of eigenvalues $\rho (\lambda)\, d\lambda$, which is
determined by a saddle point analysis at large $N$. 
Here,  $\lambda > 0$ is an eigenvalue of $M^{\dagger} M$.
For $\mu^2 > 0$ the
support of $\rho (\lambda)$ is $[0,1]$, under suitable normalisation of
$M$, and $\rho(\lambda) \sim 1/\sqrt{\lambda}$ as $\lambda \to 0$. 
For sufficiently
negative $\mu^2$ the support becomes $[A,A+1]$, $A>0$, with $\rho(\lambda)
\sim \sqrt{\lambda-A}$ as $\lambda \to A$. 
This change in the analytic structure
causes a third order phase transition in $Z$ at a certain (negative)
value of $\mu^2$, in direct analogy with the situation for Wilson 
link variables \cite{gross}. In order to avoid the phase transition,
one must ensure the support exists down to $\lambda=0$. On the other hand,
according to Section~\ref{motivate}, the effective potential should
force $M^{\dagger}M  \to 1$ as $\mu^2 \to -\infty$. These two
requirements can only be met by carefully tuning a non-polynomial effective
potential. The infrared stable trajectory would be of this sort.
The density $\rho$ would begin to vanish everywhere, except near
$\lambda=1$, 
only
as the fixed point itself is reached. Any finite-order truncation of the 
effective potential would not leave a navigable route to the fixed
point, although it may allow a trajectory at intermediate lattice
spacing close to the infrared stable trajectory. It is this latter
possibility that we have sought to establish in this paper.

\subsection{Zero modes?}

There are two sources of $k^{+} = 0$ modes in the transverse lattice
theory, both of which we have not used. Firstly the light-cone gauge
$A_{-} = 0$ is not compatible with removal of the region $k^+ = 0$,
although the similar gauge $\partial_{-} A_{-}=0$ is.
The difference is  an extra quantum mechanical degree of freedom which
will lead to a non-trivial vacuum structure in the type of 
theories considered here \cite{pinsky}. 
Very little is known about
the quantitative effects on the spectrum. Some of the effects may
just be absorbed into our effective potential. 
This dynamical zero mode certainly deserve further attention.

Secondly, there is the $k^+ = 0$ part of $M$, which we can ignore in the
dielectric regime. If we decompose $M=HU$
into a `radial' Hermitian matrix $H$ and an `angular' unitary matrix
$U$, we expect that as the mass squared of the $M$-quanta turns
negative, $H$ gets a VEV $H_0$. If $H= H_0 + \tilde{H}$, as the
lattice spacing vanishes $H_0$ freezes while $\tilde{H}$ 
becomes infinitely massive and
decouples, leaving $U$ which is nothing but the usual Wilson link
variable. This is just another way of phrasing the arguments of
Section~\ref{motivate}. 
Evidently the Hamiltonian approach becomes very complicated once
outside of the dielectric regime because of this zero mode and the 
neccesity of decomposition into radial and angular parts.
The fact that we  see a slight 
worsening of our results at very small $\mu^2$ may indicate
that we are becoming sensitive to the onset of the above 
phenomenon. Howvever we cannot rule out that these fluctuations are
due to some subtlety in the $K$ and/or $p$ extrapolations near
$\mu^2 = 0$.

\subsection{Improvements and extensions}

There are two points of view one could take with regard to fixing the
couplings in the effective potential $V[M]$. If one is mostly
interested in determining a hadron light-cone wavefunction, 
an acceptable phenomenological procedure would be to tune $V[M]$ until
it reproduces the mass spectrum from experiment or Euclidean lattice
methods. The hadronic structure functions {\em et cetera} would then be 
predictions.
We will not discuss this point of view further. Another approach is to
attempt to determine the scaling trajectory {\em a priori} without 
reference to other data by examining Lorentz invariance. 
In this paper our only measure for this was the multiplet structure of
the glueball spectrum. Although in principle one could fix $V[M]$ by
demanding the correct degeneracies, in practice we found that this was
not really tenable
since we are working at finite lattice spacing.  
Indeed we found it difficult
to get the lowest parity doublet to be degenerate.
In addition, it takes no account of
rotational invariance in the spin 0 states, which form most of light
spectrum. 
Another method is to measure the heavy
source potential in the spatial directions. In  Ref.~\cite{burkardt} this
method was studied for the same light-cone Hamiltonian problem, but
with $V[M]$ arbitrarily set to zero. Nevertheless, for distances greater 
than two lattice spacings the potential appeared remarkably 
rotationally invariant in $(x^1, x^2)$. However, over a typical hadron
size (of order two or three  lattice spacings according to our analysis)
the results were much less satisfactory. 
Also, while it can  test the scalar
part of the interaction, it is not sensitive to the full spin
dependence of the theory.

A better measure of the restoration of Lorentz invariance
is provided by the relativistic dispersion relation of each glueball 
wavefunction which allows one to keep track of Lorentz
invariance for all states, whether of zero or non-zero spin. 
To test whether the dispersion relation
$M^2 = 2P^+ P^- - (P^{1})^{2}$ holds for each glueball requires us to
work at non-zero transverse momentum $P_1$. This can be done within
the
context of the Eguchi-Kawai reduced formulation by adding appropriate
phase factors to the Hamiltonian matrix elements. It does require a
certain amount of code rewriting and we plan to include this
measurement in future work. It should enable us to fix an optimal
effective potential for symmetry restoration and thus predict the
approximate scaling trajectory without the need for a fit
to an existing spectrum.

Although we have not discussed them in any detail in this paper, perhaps
the most interesting output of our calculations are the light-cone
wavefunctions. These may be used to calculate not only distribution
functions at a given low normalisation point for the transverse momentum
cutoff, but also their non-perturbative evolution as one moves along the
scaling trajectory. A detailed analysis will be given in a future
publication once the trajectory has been better verified.

There is the possibility of performing some sort of analytic or
semi-analytic renormalisation group procedure.  As we have emphasised,
we cannot use small lattice spacing as a parameter to do this in the
dielectric regime.  However, we do have dynamical information which
suggests that mixing of Fock sectors with different numbers of
particles can be treated as a perturbation, and we can also use
perturbation in small $\mu^2$ if necessary. In this paper we have
already performed some rudimentary analytic work to find the critical
surfaces for vanishing tension and mass gap. This could probably be
extended, order by order in processes which change the number of
particles, to derive an approximate scaling trajectory at intermediate
lattice spacing.  In practice, however, it may be just as well to do
this numerically.  Basically we are dealing with a very discrete
problem in the transverse lattice direction and numerical methods are
much more efficient.  In the longitudinal directions on the other
hand, where we take the continuum limit, we have demonstrated that
semi-analytic methods produce a vast improvement in the removal of
discretisation errors.

An `equal-time' Hamiltonian investigation using colour
dielectric variables \cite{mack} would also be interesting as a 
comparison.  
However, even in the large $N$ limit,
the quantum vacuum becomes complicated, making study of the spectrum 
more difficult.

The most obvious generalization of our transverse lattice work
is to $3+1$ dimensions \cite{bard}.  The same
techniques are applicable here, although the numerical 
computation becomes more difficult. 
Although one must add more terms to the effective potential to a given
order in the fields,
there are many more physical observables which can be used
to fix the associated coupling constants.

\vspace{10mm}

\noindent
\section*{Acknowledgements} We thank M. Burkardt, Y. Kitazawa, 
H-J.\ Pirner, and M. Teper for helpful
discussions, also M. Teper for making available his ELMC results before
publication.  SD thanks Prof.\ H-C.\ Pauli (MPI Heidelberg) and
Prof.\ F. Lenz (Erlangen) for hospitality during the course of the work.
The work of BvdS  was supported in part by the Alexander Von Humboldt
Stiftung and a NATO travel grant. 
The work of SD was supported  by the Particle Physics and 
Astronomy Research Council.

\newpage
\begin{appendix}
\section{Eguchi-Kawai Reduction}
\label{appendixa}

Large-$N$ Eguchi-Kawai reduction at $P^1 = 0$ 
means that the light-cone Hamiltonian in 
the basis of Wilson loop Fock states built on $|0\rangle$ (\ref{typ}) 
is the same as the Hamiltonian
for the corresponding problem 
where the transverse lattice is compactified on $m$ 
links, $m\in \{1,2,\ldots\}$, {\em id est} one makes the identification $M_{x^1}(x^-) =
M_{x^1 + ma}(x^-)$ in both basis states and Hamiltonian. In particular
for $m=1$ this means the identification
\be
M_{x^1}(x^-) = M(x^-) \;\;\;\; \mbox{for all $x^1$.} \label{id}
\eq
For the basis states themselves the identification (\ref{id})
is obviously a $1  \to 1$
mapping since any single Wilson loop for a state of $P^1=0$ is completely
specified by its shape. The shape is completely specified by the sequence
of orientations of the link modes in the colour trace  and does not require
knowledge of the absolute positions of the modes on the lattice.

Now the original Hamiltonian (\ref{energy}) 
is quartic and local in the transverse
direction. Therefore link modes on the Wilson loop interact only if they
are within two lattice spacings of one another. Under reduction (\ref{id}),
modes which were not within two lattice spacings now lie on the same link
and will interact. Also modes which were not on the same link, but were
within two lattice spacings of one another, will have additional forms
of interaction under reduction. The problem is to show that the extra
interactions which arise from the reduction (\ref{id}) are suppressed by
$1/N$. This is a finite task since the Hamiltonian, being quartic, involves
no more than four link modes in initial plus final states. The full
proof is nevertheless quite tedious, so we illustrate the steps with a couple
of examples.

Our first example is a $2 \to 2$ amplitude which occurs on neighbouring
link modes before reduction. It is
the combination
\be
\Tr \left\{a^{\da}_{+1}(q,x^1) a^{\da}_{+1}(q^\prime,x^1+a) a_{+1}(k^\prime,x^1+a) 
a_{+1}(k,x^1) \right\} \; , \; \;\;\; q+q^\prime= k+k^\prime \; , \label{one}
\eq
which comes from the operator coupling to $\lambda_2$ in the Hamiltonian
\be
{\lambda_2 \over N} 
\Tr \left\{ M_{x^1} M_{x^1}^{\da} M_{x^1 -a}^{\da} M_{x^1-a} \right\} 
  \;  . \label{two}
\eq
On the successive link modes
\be
\Tr \left\{ \cdots a^{\da}_{+1,kj}(k^\prime,x^1+a) a^{\da}_{+1,ji}(k,x^1+a) \cdots \right\} |0\rangle 
\label{three}
\eq
the result is
\be
\lambda_2 \Tr \left\{  \cdots a^{\da}_{+1,kj}(q^\prime,x^1+a) a^{\da}_{+1,ji}(q,x^1)
\cdots \right\} |0\rangle
\eq
In the reduced theory the $x^1$ label is dropped in 
Eqns.\ (\ref{one}), (\ref{two}), and (\ref{three}). The result this time is
\begin{eqnarray}
&&\lambda_2  \Tr \left\{ 
\cdots a^{\da}_{+1,kj}(q^\prime) a^{\da}_{+1,ji}(q) \cdots \right\}|0\rangle 
\nonumber \\
&& +{\lambda_2 \over N}  \Tr \left\{ a^{\da}_{+1}(q)a^{\da}_{+1}(q^\prime) \right\}
 \Tr \left\{ \cdots \delta_{ik} \cdots \right\} |0\rangle \; .
\end{eqnarray}
The reduced amplitude picks up an extra term which is suppressed by $1/N$.
Note that the longitudinal coordinate plays no role in the calculation.

Our second example is a $3 \to 1$ amplitude involving widely separated
links before reduction.  It is the combination 
\be
\Tr \left\{ a^{\da}_{+1}(k_4,x^1) a_{+1}(k_1,x^1)a_{-1}(k_2,x^1)a_{+1}(k_3,x^1)
\right\} \ \ , \ \ k_4 = k_1 + k_2 + k_3 \ \ , 
\eq
which comes from the operator coupling to $\lambda_1$ in the Hamiltonian
\be
{\lambda_1 \over N} \Tr \left\{ M_{x^1} M^{\da}_{x^1} M_{x^1} M^{\da}_{x^1} \right\} \ . 
\label{local}
\eq
On the separated link modes
\be
\Tr \left\{ \cdots a^{\da}_{+1,ij}(k_1,x^1)a^{\da}_{-1,jk}(k_2,x^1) \ldots
a^{\da}_{+1,lm}(k_3,y^1) \cdots \right\} |0\rangle
\eq
the result is zero since (\ref{local}) is local to the link $x^1$. In the
reduced case however the result is non-zero
\be
\lambda_1 \Tr \left\{ \delta_{kl} \ldots\right\} \Tr \left\{\cdots a^{\da}_{+1,im}(k_4)
 \cdots \right\} |0\rangle
\label{sup}
\eq
Once one takes into account the $N$-normalisation of states, see
(\ref{wf}), one sees that (\ref{sup}) is $1/N$ suppressed.
The generalisation of the proof to two or more transverse dimensions
is straightforward.

Eguchi-Kawai reduction was proved for Wilson's lattice gauge theory
using the equivalence of the Dyson-Schwinger loop equations
\cite{ek}. This equivalence required that, in the reduced formulation,
extra terms proportional to the expectation value of a pair of
disconnected Wilson loops with opposite non-zero winding number
vanished. This was true under the assumption of large-$N$
factorization and unbroken $U(1)$ winding number symmetry.  However
this $U(1)$ is spontaneously broken at weak coupling ({\em id est}
small lattice spacing) \cite{twist}, spoiling the equivalence under
reduction. Our proof of equivalence in the light-cone Hamiltonian
framework based on the dielectric variables in the vacuum $|0\rangle$
makes no reference to the winding sectors at all. The extra
disconnected loops which give rise to the problems in the Dyson
Schwinger approach have no analogue here because the large-$N$
light-cone Hamiltonian propagates loops without splitting or joining
them.  This is a special feature of the light-cone quantisation which
does not hold in the equal-time quantisation.

Does this mean we avoid the breakdown of Eguchi-Kawai reduction? The
answer is yes, but unfortunately not in a way that that can take us to
small lattice spacing. Our proof is based on the dielectric variables
acting on the vacuum $|0\rangle$.  Provided this is the true vacuum,
our quantisation is valid and we can use the reduced formulation. If
the true vacuum lies elsewhere, which would be signalled by tachyons
in the spectrum (see Fig.\ \ref{fig3}), then even our quantisation of
the unreduced theory about $|0\rangle$ is not valid. In particular, our
quantisation breaks down for $m^2 < 0$, which puts a lower limit on
the lattice spacing we can achieve on the scaling trajectory; see
Fig.\ \ref{fig9}.  At present we do not know how to formulate the
light-cone Hamiltonian theory in terms of variables other than the
dielectric ones and, even then, anywhere other than about the $M=0$
classical solution. Thus, even though the light-cone formulation
avoids the usual source of breakdown of Eguchi-kawai reduction,
we are unable to exploit this to reach small
lattice spacing in the reduced theory.

\section{Tamm-Dancoff Extrapolation}
\label{appendixb}

Our numerical calculations of the glueball spectrum use a 
cutoff on the maximum number of link variables $p$ in a state,
{\em id est}
a Tamm-Dancoff cutoff on the Hilbert space. Although a popular
truncation scheme there is very little work in the light-cone 
literature on how one expects this to extrapolate to $p \to \infty$.
In the following, we will use the ``transverse only'' model of
Section~\ref{transverseonly} 
to estimate the convergence in particle truncation in the limit
of large $p$.

\begin{figure}
\centering
\BoxedEPSF{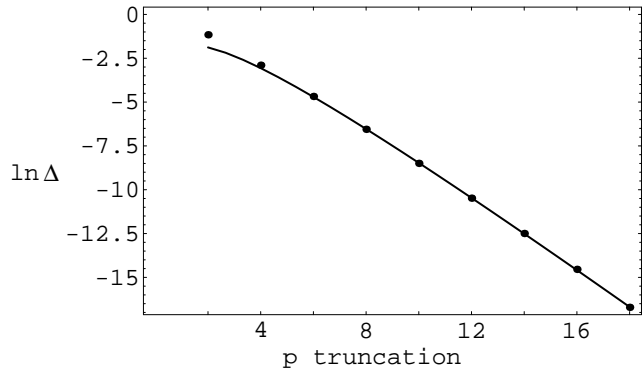 scaled 667}
\caption{ Convergence of eigenvalues 
$\ln \Delta = \ln\left|M^2_p-M^2_{\infty}\right|$ 
vs particle truncation $p$ for the ``transverse only'' model.
Here, we have chosen $\mu^2=0$,  $\lambda_1 = -0.1 g^2 N$, 
and  $\lambda_2= -0.2 g^2 N$.  The points are from a 
numerical calculation using the basis 
(\protect\ref{trbasis}). Also shown is a numerical fit to the function
  $-0.698 + c\, p + 1.597 \ln p$
  where $c = -1.145$ is given by Eqn.~(\protect\ref{cp}).
\label{fig20}}
\end{figure}

Let us symmetrise our states 
$\left| \sigma_1 , \ldots , \sigma_p \right\rangle $
under orientation reversals ${\cal  O}: \sigma_i \to \sigma_{p-i+1}$,
reflections ${\cal P}_1: \sigma_1 \to -\sigma_i$, and
cyclic permutations.  The total number of distinct states, 
after symmetrisation, 
is bounded above by $p \choose p/2$ and bounded below
by ${p \choose p/2}/(4 p)$.  For large $p$, the actual
number of distinct states should approach the lower bound
because a generic many-particle state
will not be symmetric under the above 3 operations. 
Thus, in the limit of large $p$ the number of states is
\be
  \frac{2^{p-1}}{\sqrt{2 \pi}\, p^{3/2}} 
    \label{number}
\eq
and the number of states for each value of $\gamma$ is
\be
   \frac{2^{p+1/2-4 p (\gamma-1/2)^2} \sqrt{\ln 2}}{\pi p^2}
    \; . \label{peak}
\eq
Note that the distribution becomes
increasingly peaked at $\gamma=1/2$ as $p$ is increased.
The normalisation for a generic state is
\be
   1= \left\langle \sigma_1 ,\ldots , \sigma_p \right|
 \left. \sigma_1 ,\ldots , \sigma_p \right\rangle=
  C_{\sigma_1,\ldots , \sigma_p}^2 \; .
\eq
That is, we expect only one nonzero contraction for
a generic many-particle state.
Likewise, the number of $p-2$ particle states that couple to
a given $p$ particle state is typically $p/2$.
Assuming only one nonzero contraction, the associated
matrix element is, on average,
\be
     \left\langle \sigma_1 ,\ldots , \sigma_p \right| P_\mu P^\mu
 \left| \sigma_1 ,\ldots , \sigma_{p-2} \right\rangle
    = \frac{
       p (1+2 \beta) {\Gamma{(\beta+1/2)}}^3 \Gamma(1+4 \beta)
    }{
      2 \pi a {\Gamma(1+2 \beta)}^2 \Gamma(3/2+3 \beta)
     }
     \left[2 (1-\gamma) \lambda_1 + \gamma \lambda_2 \right]
     (1+O(1/p)) \; ;
\eq
note that these two states must have approximately the 
same value of $\gamma$.
Using perturbation theory, the amplitude
for a generic $p$-particle basis state in a low energy
eigenstate is
\be
      d_p \propto \left(\frac{p 
   \left\langle \sigma_1 ,\ldots , \sigma_p \right| P_\mu P^\mu
 \left| \sigma_1 ,\ldots , \sigma_{p-2} \right\rangle
        }{2  \left\langle \sigma_1 ,\ldots , \sigma_p \right| P_\mu P^\mu
 \left| \sigma_1 ,\ldots , \sigma_p \right\rangle
   }\right)^{p/2}
\eq
times an undetermined power of $p$ (which we will ignore).
Again using perturbation theory, one finds
for a $p$-particle truncation that a low energy 
eigenvalue converges as 
\be
   M^2_p - M^2_\infty \approx
            - d_p^2 \,  2 p \, \frac{2^{p-1}}{\sqrt{2 \pi} \, p^{3/2}}
         \, \frac{{
   \left\langle  \sigma_1 ,\ldots , \sigma_p \right| 
         P_\mu P^\mu \left|  \sigma_1 ,\ldots , \sigma_{p+2}\right\rangle
    }^2 }{ 
   \left\langle  \sigma_1 ,\ldots , \sigma_{p+2} \right| P_\mu P^\mu 
          \left|  \sigma_1 ,\ldots , \sigma_{p+2}\right\rangle
} 
\eq
where we have estimated the number of states from 
Eqn.~(\ref{number}).  Since we know from Eqn.~(\ref{peak})
that the distribution of states is strongly peaked at 
$\gamma=1/2$, we set $\gamma$ to this value.  (This is most
valid when $2 \lambda_1 \approx \lambda_2$.)
Thus, the convergence in particle truncation is 
exponentially fast
\be
    M^2_p - M^2_\infty \propto e^{c \, p}
\eq
where
\be
      c = \ln \!\left|\frac{
         2 \left[\lambda_1 + \frac{\lambda_2}{2}\right] 
           {\Gamma\!\left(\beta+\frac{1}{2}\right)}^3 \,
           {\Gamma(2 \beta+1)}^2
      }{
           \Gamma\!\left(3 \beta+\frac{3}{2}\right) 
            \left(\left[g^2 N (1+4 \beta) + \lambda_1
           +\frac{\displaystyle \lambda_2}{\displaystyle 2} \right] 
               {\Gamma\!\left(\beta+\frac{1}{2}\right)}^4 
           +\frac{\displaystyle 2 \pi a \mu^2 {\Gamma(2 \beta+1)}^4 }
              {\displaystyle \beta \, \Gamma(4 \beta+1)}\right)
      }\right|
     \; .      \label{cp}
\eq
This is the formula for convergence in particle truncation that we
use in our numerical work.   As illustrated in Fig.~\ref{fig20}, 
it agrees well with numerical results for the ``transverse only''
model. 
Consequently, our biggest assumption is that it can be applied to 
the full theory as well.
%
%

\section{Improved matrix elements for DLCQ}
\label{appendixc}

In this section, we will describe a novel technique for 
improving the numerical convergence of calculations
which employ Discretised Light Cone 
Quantisation (DLCQ)~\cite{brod} to solve
the associated set of coupled integral equations.
An important advantage of DLCQ in a many particle calculation
is that the resulting Hamiltonian matrix is quite sparse and
it is essential that we preserve this property when making
any improvements.

The integer-valued ``harmonic resolution'' $K$ denotes the 
coarseness of the longitudinal momentum discretisation.
Na{\"{\i}}vely, one would expect that the leading finite $K$ 
error associated
with a DLCQ calculation would go as $1/K^2$; this is simply
the error associated with approximating an integral using the
trapezoidal rule.  However, our Hamiltonian contains several 
singularities which make the convergence substantially slower.
First, there are (integrable) endpoint singularities which go as
$1/\sqrt{k^+}$ and produce a leading error of $1/\sqrt{K}$.
Second, there is the (non-integrable) singularity associated
with the mass term and the instantaneous interaction 
$1/(k^+-p^+)^2$ which gives
a leading error of $1/K^{2 \beta}$ \cite{mass}.
Finally, the error for producing a state with more particles
goes as $1/\sqrt{K}$.

Our method is the following: when calculating the matrix elements
of the Hamiltonian in a many-particle calculation, we will use 
ordinary DLCQ commutation relations 
to calculate contractions associated with spectator particles.
To calculate the interaction itself, we take the DLCQ momenta of
the interacting particles, project onto a smooth wavefunction basis
(continuous longitudinal momentum), and calculate matrix elements
of the interaction in this smooth wavefunction basis.  
Our modification of DLCQ essentially eliminates the 
$1/\sqrt{K}$ and $1/K^{2 \beta}$ errors mentioned above.
The remaining errors include a (small) $1/K$ error from the
instantaneous interaction $1/(k^+-p^+)^2$, $1/K^2$ errors from
DLCQ itself, and a $1/K$ error from production of
particles at small momentum.

We will discuss the $(\mbox{2 particles}) \to (\mbox{2 particles})$ and 
the $(\mbox{1 particle}) \to (\mbox{3 particles})$ interactions separately
since we use somewhat different techniques for each.

\subsection{$(\mbox{2 particles}) \to (\mbox{2 particles})$ interactions}

Consider a typical 4-point interaction
%
%
\begin{equation}
V\!\left(\frac{k_1}{K},\frac{k_3}{K}\right)=
\begin{array}{r}k_1\\[25pt]k_2 \end{array}
\BoxedEPSF{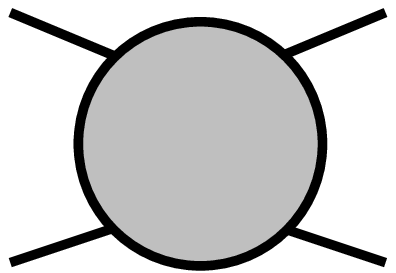 scaled 500}
\begin{array}{l}k_3\\[25pt]k_4\end{array}
     \label{vtwo}
\end{equation}
with longitudinal momenta $k_i \in \{1/2,3/2,\ldots\}$ and 
$K = k_1 +k_2 = k_3 + k_4$.
We start by calculating the first $K$ eigenstates of 
the two-particle bound state equation~\cite{bard}
\begin{eqnarray}
  M^2 \Psi(x) &=& \int_0^1 dy\,H(x,y) \, \Psi(y)\nonumber \\
  &=& \frac{\mu^2\, \Psi(x)}{x (1-x)}
    - \frac{g^2 N}{2 \pi a} \int_0^1 dy\, \frac{ \Psi(y)}{\sqrt{x (1-x) y
   (1-y)}}\left(\frac{(x+y)(2-x-y)}{(x-y)^2} +1\right)
  \label{twoe}
\end{eqnarray}
where a principal value prescription is assumed for the integral.
Here, we have chosen 
$2 \lambda_1 +\lambda_2+2 \lambda_3 = - g^2 N$ 
so that, at $\mu^2=0$, the lowest wavefunction is constant with
eigenvalue zero.  Then we write the incoming and outgoing DLCQ states 
as linear combinations
of these $K$ eigenfunctions and evaluate matrix elements of the 
interaction $V(x,y)$ in the eigenfunction basis.  The result is expressed
as a Taylor expansion in $\beta$.  
Details are described in the following.

We introduce a basis of even and odd polynomial wavefunctions
(labeled by subscripts $s$ and $s^\prime$):
\begin{equation}
  \phi_s\!(x) = (x(1-x))^{\beta+s-1} \left\{\begin{array}{c}1\\1-2x\end{array}
               \right. \; ,
\end{equation}
$s \in \{1,2,\ldots,S\}$, $S \gg K/2$, and $\beta$ 
was defined in Eqn.~(\ref{vanish}).
We define, for both even and odd wavefuntions, 
matrix elements of the interaction term  
$V_{s,s^\prime} = \int dx\, dy \, \phi_s(x)\, V\!(x,y)\, \phi_{s^\prime}(y)$, 
the inner product matrix 
$E_{s,s^\prime} = \int dx\, dy \, \phi_s(x)\, \phi_{s^\prime}(y)$, and
the two-particle Hamiltonian (\ref{twoe})
$H_{s,s^\prime} = \int dx\, dy \, \phi_s(x)\, H(x,y)\, \phi_{s^\prime}(y)$.  
We can Taylor expand these matrices in powers of $\beta$: 
$V_{s,s^\prime}=V^{(0)}_{s,s^\prime}+\beta V^{(1)}_{s,s^\prime}+ \cdots$,
$E_{s,s^\prime}=E^{(0)}_{s,s^\prime}+\beta E^{(1)}_{s,s^\prime}+ \cdots$, and
$H_{s,s^\prime}=H^{(0)}_{s,s^\prime}+\beta H^{(1)}_{s,s^\prime}+ \cdots$.
Using the formulas in Refs.~\cite{bard,harada} along with the
definition of the Beta function 
$B(\alpha,\beta)=\int_0^1 dx\, x^{\alpha-1} (1-x)^{\beta-1}$, 
$V^{(n)}_{s,s^\prime}$, $E^{(n)}_{s,s^\prime}$, and  $H^{(n)}_{s,s^\prime}$ 
can be determined analytically.
In addition, we define a $K\times S$ matrix $T_{i,s}$ which maps the 
wavefunction basis onto the DLCQ basis (the subscript 
$i \in \{1,2,\ldots\,K\}$
denotes a two particle DLCQ basis state with momenta 
$k_1=i-1/2$, $k_2=K-k_1$)
\begin{equation}
  T_{i,s} = T^{(0)}_{i,s}+\beta T^{(1)}_{i,s}+\cdots =
         \left(\frac{k_1 k_2}{K^2}\right)^{
            \beta+s-1}\left\{\begin{array}{c}1 \\ \frac{k_1-k2}{K}
             \end{array}\right. \; .
\end{equation}
However, the polynomial basis that we introduced has a fatal
flaw: the inner product matrix $E_{s,s^\prime}$ is very poorly 
conditioned~\cite{numerical}.  
Instead, we find a new, well-conditioned, basis (denoted by subscripts
$\nu$, $\rho$, and $\sigma$) in which $E^{(0)}$ is 
diagonal.  Thus, for both even and odd wavefunctions, 
we define the Cholesky decomposition~\cite{numerical}\footnote{
The matrix elements of the lower triangular matrix $L^{-1}$ 
are related to the coefficients of the Legendre polynomials
defined on the interval [0,1].} 
$E^{(0)}=L L^T$ and calculate the matrix products
\begin{eqnarray}
V^{(n)}_{\nu,\rho}&=&\left(L^{-1}\right)_{\nu,s} V^{(n)}_{s,s^\prime} 
  \left(L^{-T}\right)_{s^\prime,\rho} \; , \label{matrixproducts1}\\
E^{(n)}_{\nu,\rho}&=&\left(L^{-1}\right)_{\nu,s} E^{(n)}_{s,s^\prime} 
  \left(L^{-T}\right)_{s^\prime,\rho} \; , \\
H^{(n)}_{\nu,\rho}&=&\left(L^{-1}\right)_{\nu,s} H^{(n)}_{s,s^\prime} 
  \left(L^{-T}\right)_{s^\prime,\rho} \; , \\
T^{(n)}_{i,\nu}&=&T^{(n)}_{i,s} \left(L^{-T}\right)_{s,\nu} \; . 
    \label{matrixproducts2} 
\end{eqnarray}
Because the polynomial basis is poorly conditioned, 
all of the above calculations, 
including the matrix products (\ref{matrixproducts1}) -- 
(\ref{matrixproducts2}), are 
numerically unstable and must be performed either exactly or 
using high precision arithmetic.

Working in the well-conditioned basis, 
we solve generalised eigenvalue problem
$H \chi_t = h_t E \chi_t$ with eigenvalues 
$h_t = h^{(0)}_t+ \beta h^{(1)}_t+\cdots$ and eigenvectors
$\chi_t = \chi^{(0)}_t+ \beta \chi^{(1)}_t+\cdots$.
First, we solve the leading order eigenvalue problem $H^{(0)} \chi^{(0)}_t = 
h^{(0)}_t \chi^{(0)}_t$ (recall $E^{(0)}_{\nu,\rho}$ is the 
identity).
Nonleading powers in $\beta$ are calculated 
using Rayleigh-Schr\"{o}dinger perturbation theory extended to
the generalised eigenvalue problem.
Thus, if we write
\begin{equation}
    \chi_t = \chi^{(0)}_t+ \beta \sum_{u} c^{(1)}_{t,u} \chi^{(0)}_u
             + \cdots 
\end{equation}
then
\begin{eqnarray}
    h^{(1)}_t &=& {\chi^{(0)}_t}^{\dagger} \left(H^{(1)}-h^{(0)}_t E^{(1)}
                  \right) \chi^{(0)}_t \\
    c^{(1)}_{t,u} &=& \frac{{\chi^{(0)}_u}^{\dagger} 
                \left(H^{(1)}-h^{(0)}_t E^{(1)}\right) \chi^{(0)}_t}
                {h^{(0)}_t-h^{(0)}_u}\; , \;\;\;\; t \neq u \\
    c^{(1)}_{t,t} &=& -\frac{1}{2} \, {\chi^{(0)}_t}^{\dagger} E^{(1)}
                       \chi^{(0)}_t  \\
              &\vdots& \nonumber
                    \; .
\end{eqnarray}
Using the lowest $K$ eigenvectors, 
$\chi_t$, we define an eigenstate basis
(denoted by subscripts $t$ and $t^\prime$),
\begin{equation}
T_{i,t}=T_{i,\nu} \left(\chi_t\right)_{\nu} \;  .
\end{equation}

The matrix $T_{i,t}$ maps the eigenfunction basis onto the DLCQ
basis and $T_{i,t} T_{t,j}$ is nearly equal to the identity.
In order that the map between the eigenfunction basis
and the DLCQ basis preserves eigenvalues, we
find an orthonormal matrix that is nearly equal to $T_{i,t}$.
Thus, we perform a standard QR-factorisation~\cite{numerical}\footnote{That
is, Gram-Schmidt orthogonalisation.} $T=QR^T$ where 
$Q=Q^{(0)}+\beta Q^{(1)}+\cdots$ is orthogonal and 
$R^T={R^{(0)}}^T+\beta {R^{(1)}}^T+\cdots$ is upper triangular.  
To obtain non-leading orders in $\beta$, we use the relation
\begin{equation}
   {Q^{(0)}}^T Q^{(n)} + {R^{(n)}}^T {R^{(0)}}^{-T} =
   {Q^{(0)}}^T \left(T^{(n)}-\sum_{m=1}^{n-1} Q^{(m)} {R^{(n-m)}}^T \right) 
       {R^{(0)}}^{-T} \label{qr1}
\end{equation}
and the orthonormality condition
\begin{equation}
   {Q^{(n)}}^T Q^{(0)} + {Q^{(0)}}^T Q^{(n)} =
        -\sum_{m=1}^{n-1} {Q^{(m)}}^T Q^{(n-m)} \; .
       \label{qr2}
\end{equation}
Eqn.~(\ref{qr2}) determines the diagonal matrix elements of 
${Q^{(0)}}^T Q^{(n)}$. We can use (\ref{qr1}) to determine
the strictly lower triangle of ${Q^{(0)}}^T Q^{(n)}$ and 
then we use (\ref{qr2}) to determine the upper triangle of
${Q^{(0)}}^T Q^{(n)}$. Finally, we use (\ref{qr1}) to determine
${R^{(n)}}^T {R^{(0)}}^{-T}$.  From this, we can
determine $Q^{(n)}$ and ${R^{(n)}}^T$.

\begin{figure}
\centering
\begin{tabular}{@{}c@{}c@{}}
\BoxedEPSF{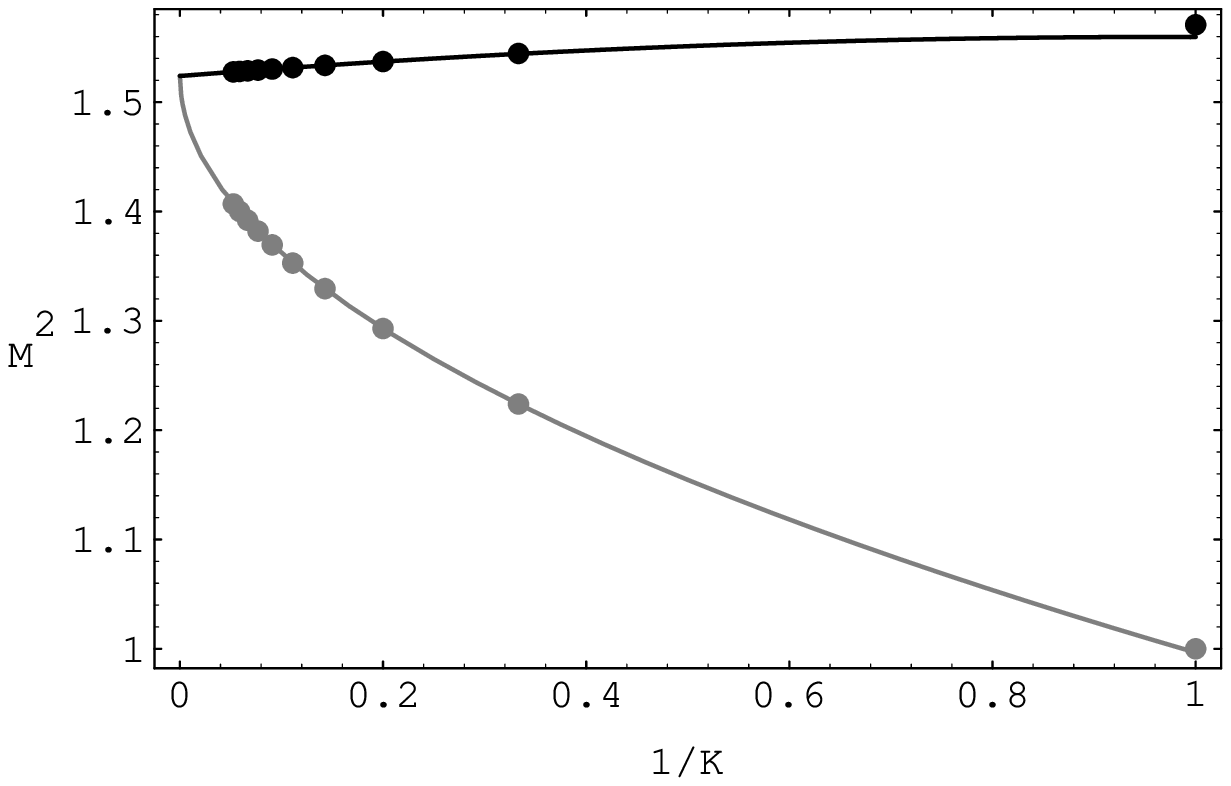 scaled 590}&
\BoxedEPSF{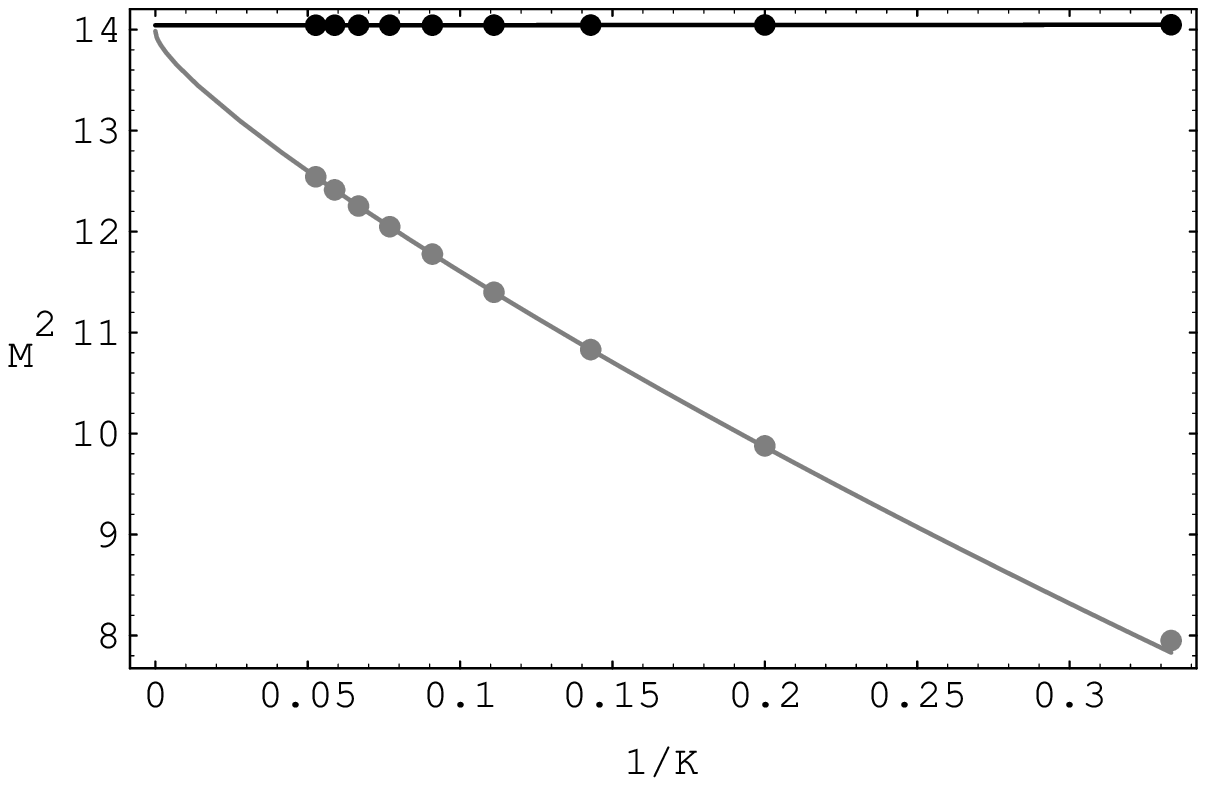 scaled 590}\\
 (a) & (b)
\end{tabular}
\caption{Convergence of the (a) ground and (b) second excited states 
vs $1/K$, two particle truncation, 
$\lambda_1 = \lambda_2 = \lambda_3 = 0$ for $\mu^2=0$.  
The grey points are from standard DLCQ and
the black points are from using the improved matrix elements.  In (b),
the eigenvalue for the improved matrix elements is 
$M^2 \approx 14.043 + 0.012/K + 0.0085/K^2$.
\label{twoconv}}
\end{figure}

\begin{figure}
\centering
\begin{tabular}{@{}c@{}c@{}}
\BoxedEPSF{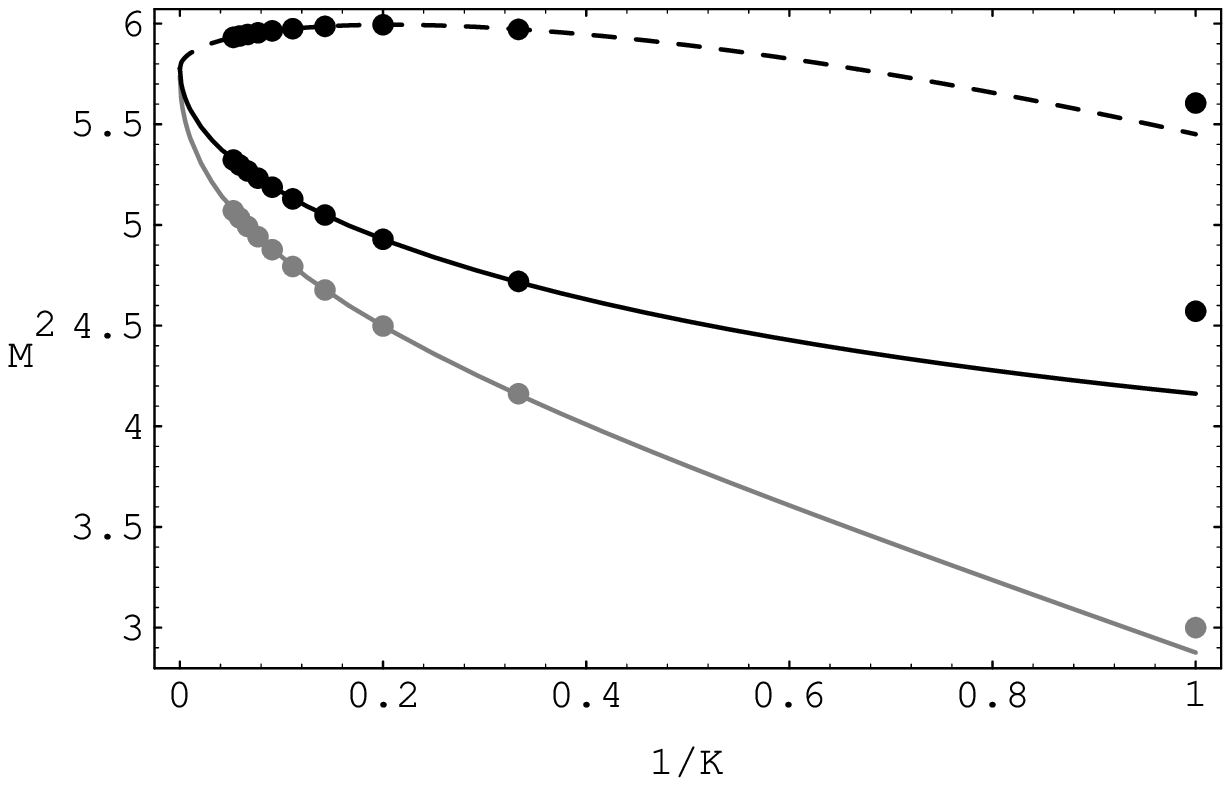 scaled 590}&
\BoxedEPSF{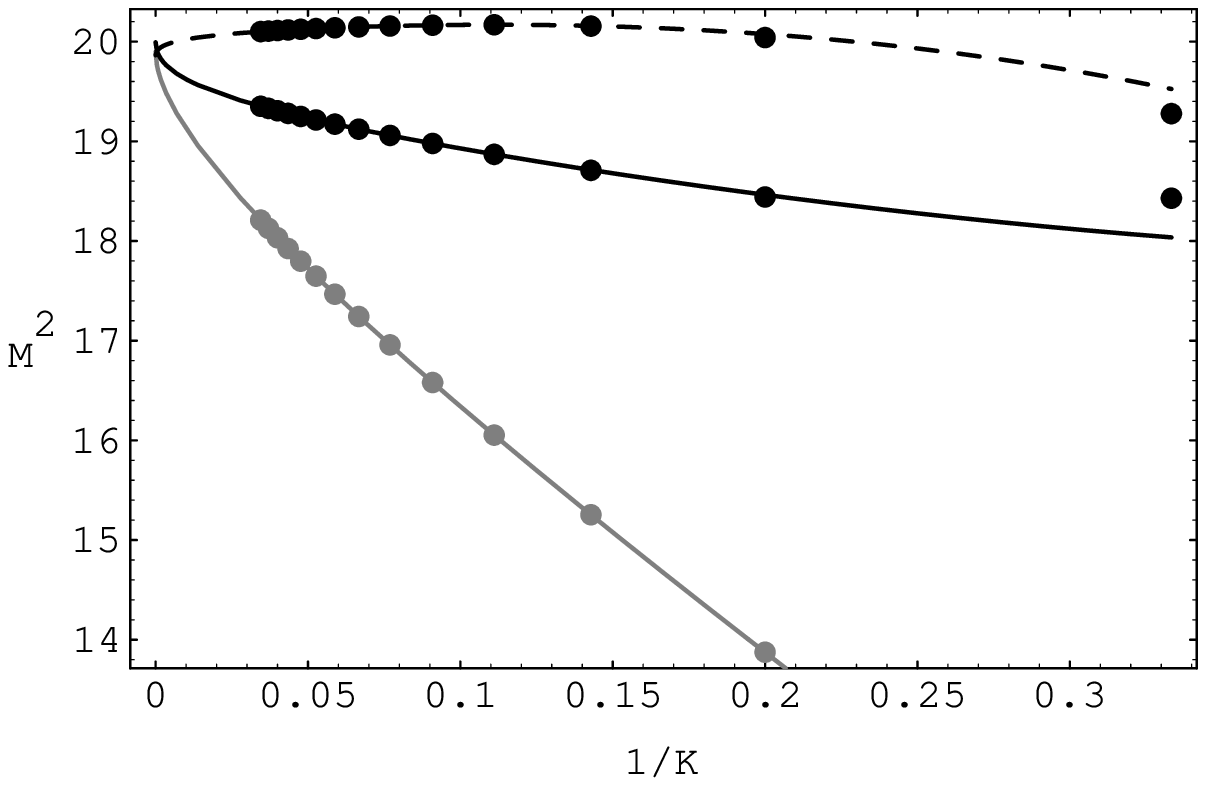 scaled 590}\\
 (a) & (b)
\end{tabular}
\caption{Convergence of the (a) ground and (b) second excited states 
vs $1/K$, two particle truncation, 
$\lambda_1 = \lambda_2 = \lambda_3 = 0$ for $\mu^2=g^2 N/(2 a)$ or
$\beta=1/4$.  
The grey points are from standard DLCQ, the 
black line is with 
the improved matrix elements expanded to order $\beta^0$, and
the dashed line is with 
the improved matrix elements expanded to order $\beta^1$.
\label{twoconv2}}
\end{figure}

We discard $R^T$ and use $Q$ to rotate to the DLCQ basis,
\begin{equation}
    V_{i,j} = Q_{i,t} \left(\chi_t\right)^\dagger_\nu
            V_{\nu,\rho} \left(\chi_{t^\prime}\right)_\rho Q^T_{t^\prime,j}
          \; .
\end{equation}
$V_{i,j}$ can be calculated order-by-order in $\beta$. 
To see how well the improved matrix elements work, we can
look at the eigenvalues vs $K$ for a two particle truncation.
The convergence with $K$ for $\mu^2=0$ and $\mu^2 = g^2 N/(2 a)$
is shown in Figs.~\ref{twoconv} and \ref{twoconv2}.

\subsection{$(\mbox{1 particle}) \to (\mbox{3 particles})$ interactions}

Consider a typical 4-point interaction
%
%
\begin{equation}
V\!\left(\frac{k_1}{K},\frac{k_2}{K},\frac{k_3}{K}\right) =
 \; K \BoxedEPSF{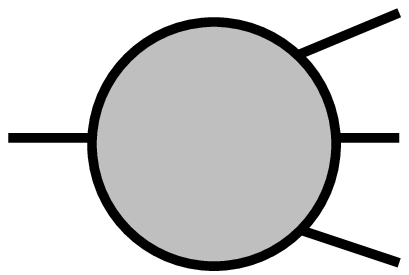 scaled 500}
\begin{array}{c}k_1\\[10pt] k_2 \\[10pt] k_3
\end{array} 
\label{vthree}
\end{equation}
where $k_i \in \{1/2,3/2,\ldots\}$ and $K = k_1 +k_2 + k_3$.
\begin{figure}
\centering
\BoxedEPSF{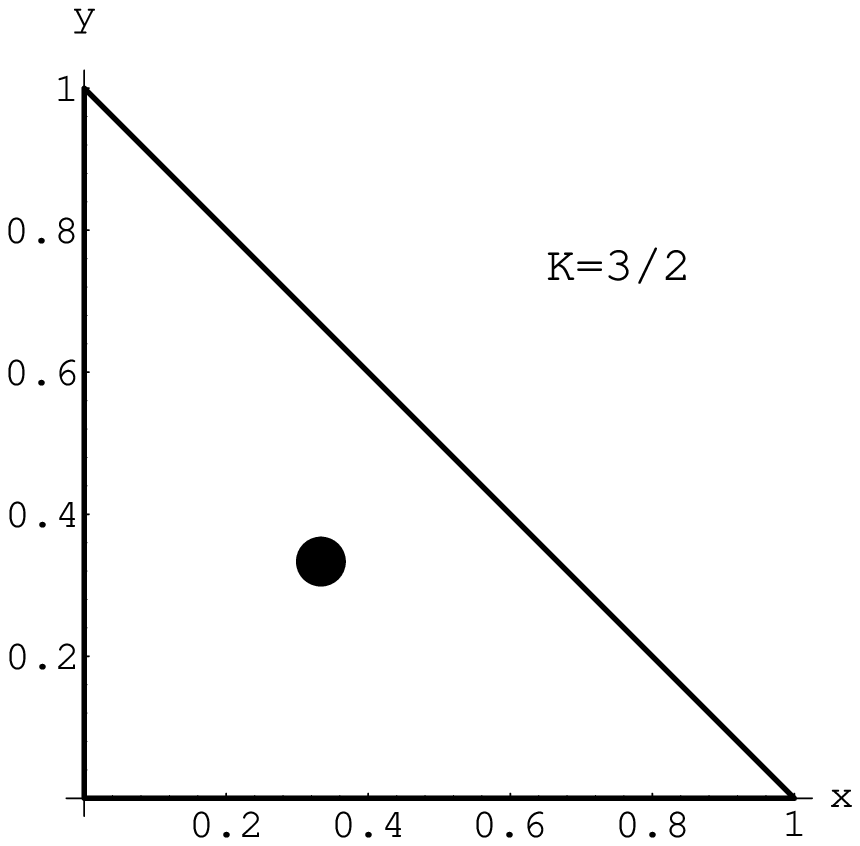 scaled 550}\hfill
\BoxedEPSF{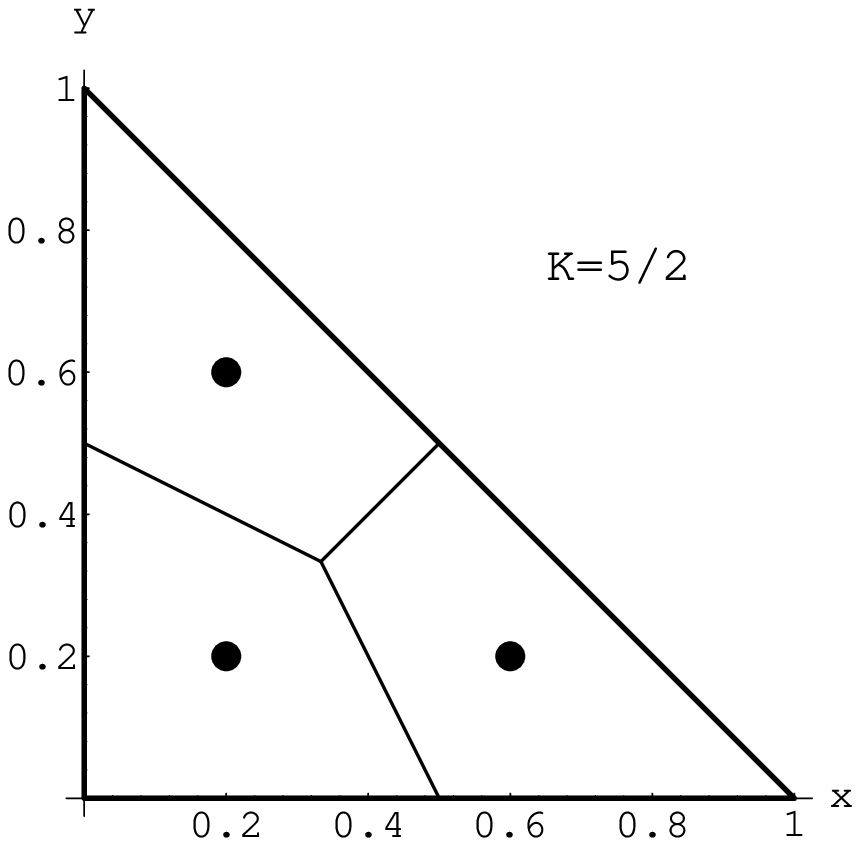 scaled 550}\hfill
\BoxedEPSF{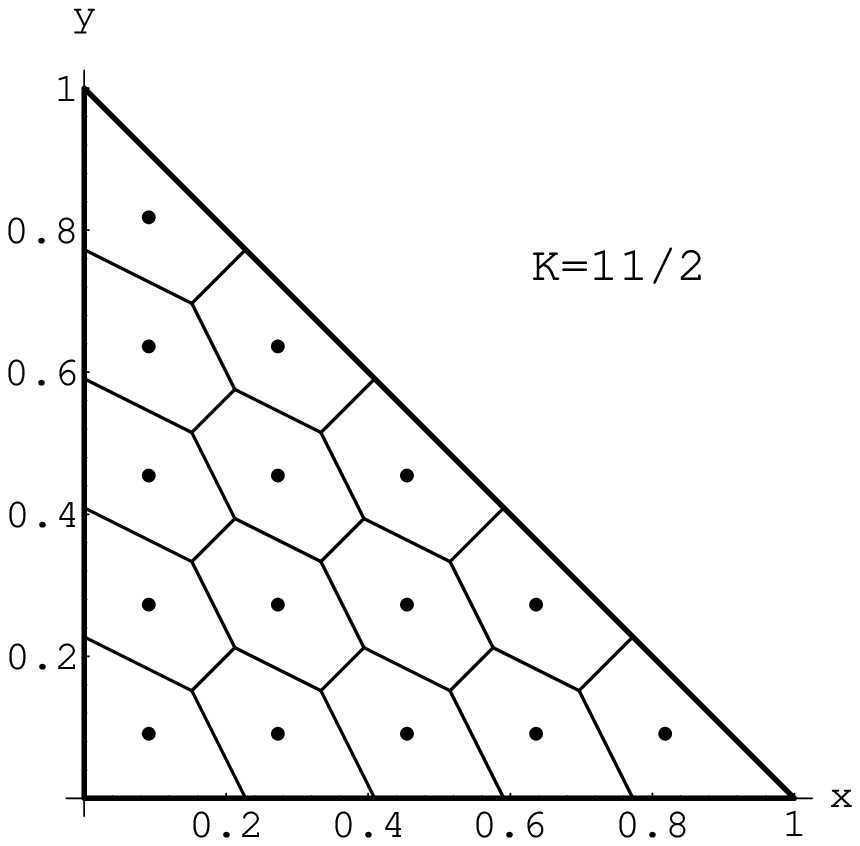 scaled 550}
\caption{Region $R_{k_1,k_2,k_3}$ for $K=3/2$, $K=5/2$, and $K=11/2$.
  The dots represent the points $(k_1/K,k_2/K)$ and $R_{k_1,k_2,k_3}$
    is the surrounding polygonal region.
   \label{triangles}}
\end{figure}
We introduce a two-dimensional space of continuous longitudinal
momentum fractions; if the three outgoing particles have
longitudinal momentum fractions 
$x$, $y$, and $1-x-y$, the corresponding  point in this space is 
labeled by $(x,y)$.  We define a polygonal region $R_{k_1,k_2,k_3}$ 
surrounding $(k_1/K,k_2/K)$ as shown in Fig.~\ref{triangles}.
Next we define a basis of wavefunctions
\begin{equation}
    \phi_{k_1,k_2,k_3}(x,y)= \left\{\begin{array}{cl}
            C_{k_1,k_2,k_3} \left(x y (1-x-y)\right)^\beta 
           \; , & (x,y)\in R_{k_1,k_2,k_3} \\
           0 \; , & \mbox{otherwise}
         \end{array} \right.
\end{equation}
where the normalisation constant $C_{k_1,k_2,k_3}$ is given by
\begin{equation}
    1= \int_0^1 dx \int_0^{1-x} dy \, {\phi_{k_1,k_2,k_3}(x,y)}^2
     \; . 
\end{equation}

We calculate improved matrix elements associated with
interaction $V(x,y,1-x-y)$ using the expression
\begin{equation}
  V_{k_1,k_2,k_3} =  \int_0^1 dx \int_0^{1-x} dy\, \phi_{k_1,k_2,k_3}(x,y)
                      V(x,y,1-x-y) \; .
\end{equation}
$V_{k_1,k_2,k_3}$ is calculated numerically order-by-order in $\beta$.

\subsection{Implementation}

In our numerical work, we expand the interaction terms to order $\beta^1$,
introducing an order $\beta^2$ error.
Thus, for large $\mu^2$, convergence with $K$ is somewhat degraded 
although the eigenvalues do still converge correctly.  
{}From Fig.~\ref{twoconv2}, we see that this is not a significant problem 
if $\mu^2<0.5 g^2 N/a$.

Even though the mass term is a two-point interaction, we 
calculate its matrix elements as if it were a four-point interaction.
Note that the expansion of the mass term in powers of $\beta$, 
has an order $1/\beta$ contribution.

In the $(\mbox{2 particles}) \to (\mbox{2 particles})$ case, we 
use $S=40$ states in our polynomial wavefunction basis; 
calculations in this basis were performed with Mathematica using
exact arithmetic (except for $T_{i,s}$ where high-precision
floating point arithmetic was used).  
For, the $(\mbox{1 particle}) \to (\mbox{3 particles})$ calculations,
we used standard numerical techniques to perform the associated 
integrations.

In practice, we calculate matrix elements of the various interaction
terms 
beforehand and store them in a file.  In a many-particle calculation,
the computer program simply looks up the appropriate matrix element 
in a table~\cite{www}.

\end{appendix}

\end{document}